\documentclass[twocolumn,showpacs,floatfix,longbibliography]{revtex4-1}
\usepackage{graphicx}
\usepackage{bm} % bold math
\usepackage{amssymb} % use this package to enable \nrightarrow command
\usepackage{amsmath} % use this package to enable \xrightarrow command
\usepackage{braket} % use for Dirac bra-kets : \rangle \labgle & \mid
\usepackage{natbib} % bibtex package
\usepackage{hyperref}

\begin{document}

\title{Currents Induced by Magnetic Impurities in Superconductors \\* with Spin-Orbit Coupling}

\author{Sergey S. Pershoguba$^{1}$}
\author{Kristofer Bj\"{o}rnson$^2$}
\author{Annica M. Black-Schaffer$^2$}
\author{Alexander V. Balatsky$^{1,3}$}

\affiliation{$^1$Nordita, Center for Quantum Materials, KTH Royal Institute of Technology, and Stockholm University, Roslagstullsbacken 23, S-106 91 Stockholm, Sweden}
\affiliation{$^2$Department of Physics and Astronomy, Uppsala University, Box 516, S-751 20 Uppsala, Sweden}
\affiliation{$^3$Institute for Materials Science, Los Alamos National Laboratory, Los Alamos, New Mexico 87545, USA}

\date{ \today}
% version P, edited by SP on June 23rd 

\begin{abstract}
We show that superconducting currents are generated around magnetic impurities and ferromagnetic islands proximity coupled to superconductors with finite spin-orbit coupling. Using the Ginzburg-Landau theory, T-matrix calculation, as well as self-consistent numerical simulation on a lattice, we find a strong dependence of the current on the direction and magnitude of the magnetic moment. We establish that in the case of point magnetic impurities, the current is carried by the induced Yu-Shiba-Rusinov (YSR) subgap states. In the vicinity of the phase transition, where the YSR states cross at zero energy, the current increases dramatically. Furthermore, we show that the currents are orthogonal to the local spin polarization and, thus, can be probed by measuring the spin-polarized local density of states. 
\end{abstract}


\maketitle

Superconductor-ferromagnet heterostructures were recently proposed as a viable platform for realizing topological superconductivity (TS) \cite{Lutchyn2010,Oreg2010, Sau2010}, which can host Majorana fermion quasiparticles at vortex cores and boundaries \cite{Kitaev2001, Alicea, Beenakker2013}. The Majorana fermions obey non-Abelian statistics and may be utilized for topological quantum computation \cite{Read2000, Ivanov2001, Nayak2008}.  The key ingredients driving these systems into the topologically nontrivial regime are the spin-orbit coupling (SOC) and magnetism. Recently, the search for experimental realizations of TS has also led to engineering the Yu-Shiba-Rusinov (YSR) \cite{Yu,Shiba,Rusinov} bands induced by magnetic atoms on the surface of a superconductor \cite{Choy2011, Nadj-Perge2013, Klinovaja2013, Vazifeh2013, Braunecker2013, Pientka2013, Nakosai2013, Poyhonen2014, Kim2014a, Reis2014, Brydon2015, Rontynen2014, Li2015}. Following this recipe, zero-energy peaks in the tunneling spectrum were recently measured at the ends of a one-dimensional (1D) chain of magnetic atoms \cite{Yazdani2014}. Such a tunneling spectrum could be the evidence of the Majorana edge states, although alternative explanations are also possible \cite{Sau2015}.

\begin{figure} \centering
(a) \includegraphics[width=0.5\linewidth]{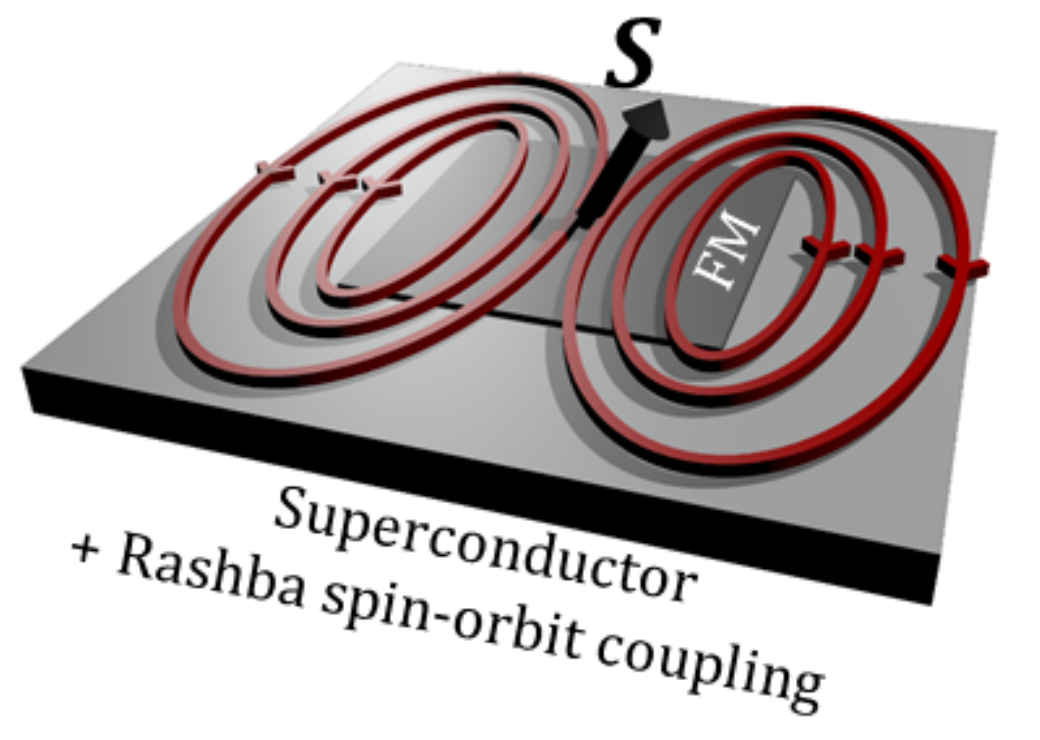}  \\
(b) \includegraphics[width=0.4\linewidth]{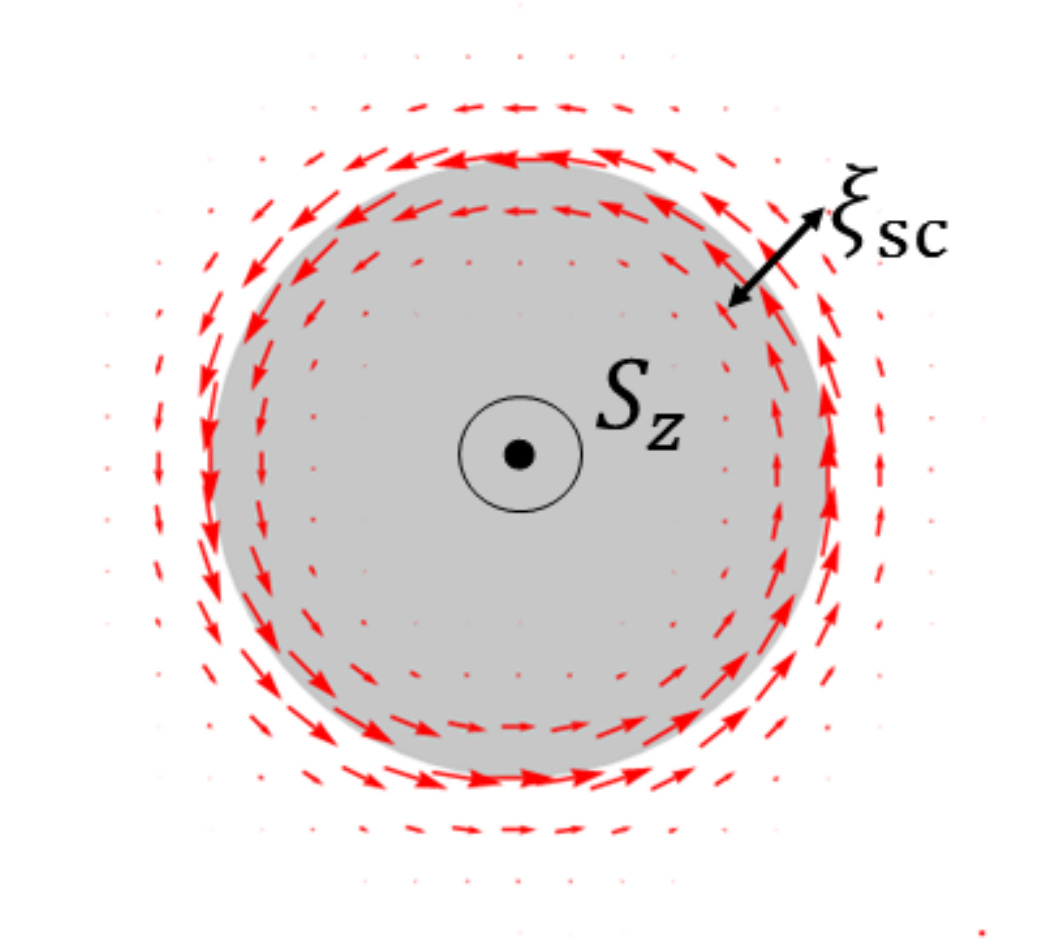} 
(c) \includegraphics[width=0.4\linewidth]{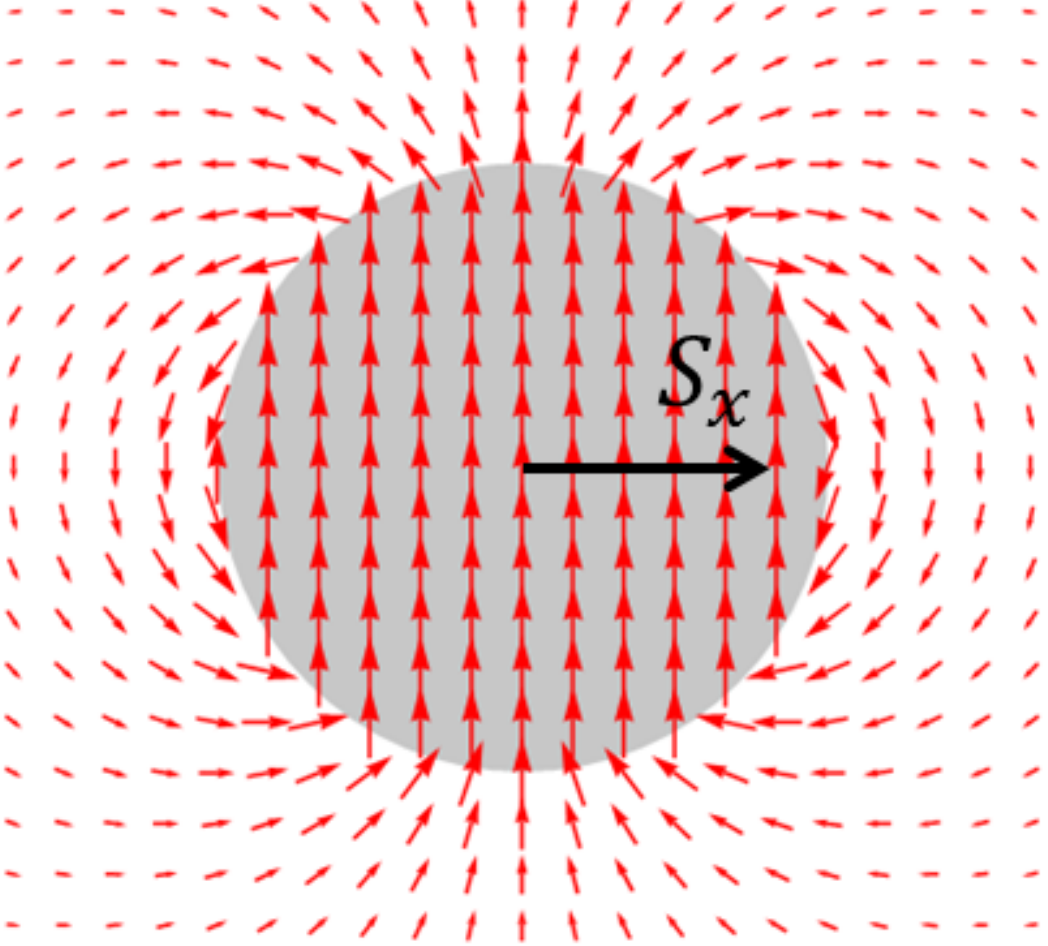}
\caption{(color online) (a) Schematic representation of the nonlocal currents (red arrows) induced by a ferromagnetic (FM) island on the surface of a superconductor with the Rashba SOC. GL solutions for the current around a circular ferromagnetic island (gray area) with $\bm S = S\, \hat{\bm z}$ (b) and $\bm S = S\,\hat{\bm x}$ (c). } \label{fig1}

\end{figure}

The interplay of SOC and magnetism has another remarkable consequence. Consider a two-dimensional (2D) surface of a three-dimensional (3D) material. The effective Hamiltonian of the surface $h(\bm p) = \frac{{\bm p}^2}{2m} + \lambda\left( \bm\sigma\times\bm p\right)_z$ contains the Rashba SOC due to the absence of the inversion symmetry at the surface. Then, the velocity operator $\bm v= {\frac{dh(\bm p)}{d\bm p} =\frac{\bm p}{m}+ \lambda \,\hat{\bm z}\times\bm\sigma}$ contains a spin-dependent term that gives an extra contribution to the current
\begin{equation}
	\bm j_{\rm extra} = \lambda\,  \hat{\bm z}\times\langle\bm\sigma\rangle . \label{main}
\end{equation}
A ferromagnet proximity coupled to the superconductor would render a finite spin polarization $\langle \bm \sigma \rangle \neq 0$ and, thus, generate a current as schematically shown in Fig.~\ref{fig1}(a). The phenomenon of driving a current with magnetism is known as the magnetoelectric effect. This effect may vanish in metals due to dissipation but survives in superconductors lacking inversion symmetry \cite{Levitov1985, Edelstein1989, Edelstein1995, Yip2001, BauerSigrist2012}. The magnetoelectric effect was also recently discussed in a pure 1D model of TS \cite{Ojanen2012}. 

%%%%%%%%%%%%%%%%%%%%%%%%%%%%%%%%%%%%%%%%%%%%%%%%%%%%%%%%%%%%%%%%%%%%%%%%%%%
%%%%%%%%%%%%%%%%%%%%%%%%%%%%%%%%%%%%%%%%%%%%%%%%%%%%%%%%%%%%%%%%%%%%%%%%%%
In this work, we show that the magnetoelectric current is universally generated around single magnetic impurities and ferromagnetic islands, which have been recently studied in the context of TS \cite{Choy2011, Nadj-Perge2013, Klinovaja2013, Vazifeh2013, Braunecker2013, Pientka2013, Nakosai2013, Poyhonen2014, Kim2014a, Reis2014, Yazdani2014, Brydon2015, Rontynen2014, Li2015}. More specifically, we first derive the extra terms in the Ginzburg-Landau (GL) free energy corresponding to Eq.~(\ref{main}). For a small ferromagnetic island on a superconductor with SOC, we find a strong dependence of the current on the relative orientation of the ferromagnetic moment. The current circulates around the ferromagnetic island and is short ranged for the ferromagnetic moment normal to the surface. On the other hand, the current has a dipolar power law decay for the ferromagnetic moment parallel to the surface. Next, we discuss the  current generated around a point magnetic impurity and show that the current is carried by the impurity-induced YSR states. We also perform a self-consistent numerical calculation and find a strong nonmonotonic dependence of the current on the strength of the ferromagnetic moment. The current strongly peaks at the phase transition, where the YSR states cross zero energy $E=0$. We further demonstrate that the current can be mapped by measuring the spin-polarized local density of states (SP LDOS), which, thus, provides a probe of both the current and the phase transition. Our findings are, therefore, highly relevant for the ongoing search of the Majorana bound states in ferromagnetic chains \cite{Choy2011, Nadj-Perge2013, Klinovaja2013, Vazifeh2013, Braunecker2013, Pientka2013, Nakosai2013, Poyhonen2014, Kim2014a, Reis2014, Yazdani2014, Brydon2015, Rontynen2014, Li2015}.

%%%%%%%%%%%%%%%%%%%%%%%%%%%%%%%%%%%%%%%%%%%%%%%%%%%%%%%%%%%%%%%%%%%%%%%%%%%
\paragraph*{Ginzburg-Landau treatment.-} \label{sec:GL}
%%%%%%%%%%%%%%%%%%%%%%%%%%%%%%%%%%%%%%%%%%%%%%%%%%%%%%%%%%%%%%%%%%%%%%%%%%
We start by considering a ferromagnetic island deposited on a 2D surface of a conventional $s$-wave superconductor with the Rashba SOC as illustrated in Fig.~\ref{fig1}(a) and described by the Hamiltonian
\begin{align}
	H & = \frac{1}{2}\int d^2\bm r\,\Psi^\dag(\bm r)\left[ h(\bm p)\, \tau_z+\Delta\, \tau_x-\bm S(\bm r) \cdot \bm \sigma\right]\,\Psi(\bm r), \nonumber  \\
	 & h(\bm p) =\frac{p^2}{2m}+\lambda\,\left( \bm\sigma\times \bm p \right)_z  - \mu,\quad \bm p = -i(\nabla_x,\nabla_y). 	\label{H}
\end{align}
Here $\Psi = (\psi_\uparrow, \psi_\downarrow, \psi^\dagger_\downarrow, -\psi^\dagger_\uparrow)^{\rm T}$ is a four component spinor, $\sigma$ and $\tau$ are the Pauli matrices acting in the spin and particle-hole Nambu space, $\Delta$ is the superconducting gap, and we set $e=\hbar=1$. The ferromagnet and its coupling to the superconductor are described by the spatially dependent vector $\bm S=(S_x,S_y,S_z)$.  
An intuitive and qualitatively correct picture of the currents can be derived using the GL free energy 
\begin{align}
	F &= \! \int \! d^2r\left[ \frac{n_s}{2m}\bm {\mathcal A}^2 +\alpha\, (\hat{\bm z}\times \bm S ) \! \cdot \! \bm{\mathcal A}+\beta\, (\bm \nabla S_z\times \hat{\bm z}) \! \cdot \! \bm{\mathcal A}  \right], \nonumber\\
	\bm {\mathcal A} &=(\mathcal A_x,\mathcal A_y) = \bm A+\frac{\bm \nabla \theta}{2}, \label{GL}
\end{align}
which is valid at length scales larger than the superconducting coherence length $\xi_{\rm sc}$. In the first term proportional to the superfluid density $n_s$, vector $\bm{\mathcal A}$ encapsulates both the superconducting phase $\theta$ and the vector potential $\bm A$. The second and third terms describe the coupling between the Rashba SOC and magnetism and are derived in the appendix. For example, in the limit $p_F\lambda\gg m\lambda^2\gg \Delta$, the coefficients are \footnote{Note that we do the calculation at $T=0$, where the magnetoelectric coefficients have a discontinuous jump between the normal and superconducting phases. In contrast at finite temperature $T\neq 0$, the magnetoelectric coefficients interpolate smoothly as $\propto \Delta^2$ between the two phases} $\alpha = m\lambda/2\pi$, $\beta=m^2\lambda^2/4\pi p_F^2$, and, thus, only present at finite SOC, i.e. when $\lambda \neq 0$. The term proportional to $\alpha$, known as the magnetoelectric term \cite{Levitov1985,Edelstein1989,Edelstein1995,BauerSigrist2012,Samokhin2004,Malshukov2014}, is  allowed  only in the absence of inversion symmetry. 

Within the above framework, we now discuss the currents induced by a ferromagnetic island of a uniform disc geometry, which we model as ${\bm S(\bm r) = \bm S \, \theta_H \left( R-r \right)}$, where $\theta_H(z)$ is the Heaviside theta function. We find the current from Eq.~(\ref{GL}) as
\begin{equation}
	\bm j = \left.\frac{\delta F}{\delta \bm A}\right|_{\bm A = 0}\! = \frac{n_s}{2m}\bm\nabla\theta+\alpha(\hat{\bm z}\times \bm S )+\beta (\bm \nabla S_z\times \hat{\bm z}). \label{current}
\end{equation}
First consider the an out-of-plane ferromagnetic moment $\bm S = S\,\hat{\bm z}$ and $\theta$ constant. Then the current is given by the last term in Eq.~(\ref{current}). The current is localized near the boundary as $\bm j(\bm r) = -\beta S (\hat{\bm r}\times \hat{\bm z})\,\delta(r-R)$ and circulates around the ferromagnetic island as shown in Fig.~\ref{fig1}(b). Since the GL equations are valid at $r>\xi_{sc}$,  the $\delta$ function in the current solution is artificially broadened to a scale of the superconducting coherence length $\xi_{\rm sc}$ for visualization purposes.  For an in-plane moment $\bm S = S\,\hat{\bm x}$, both of the first two terms in Eq.~(\ref{current}) are nonzero. The contribution given by the term $\alpha(\hat{\bm z}\times \bm S)$ is constant over the region covered by the ferromagnetic region and discontinuous at the boundary. However, the first term $\frac{n_s}{2m}\bm\nabla\theta$ fixes this discontinuity. Indeed, the variation of the free energy over $\theta$ gives the continuity equation: $0= \bm \nabla \cdot \bm j =   \frac{n_s}{2m} \bm \nabla^2\theta+\alpha\,\bm\nabla\cdot(\hat{\bm z}\times \bm S )$. The last expression is the 2D Poisson equation with a source term that we solve for $\theta$ and plot the currents in Fig.~\ref{fig1}(c), see appendix for more details. The current is constant over the region covered by the ferromagnet, i.e. $\bm j(\bm r)=\frac{\bm d}{R^2}$ for $r < R$, and has a dipolar profile outside of it, i.e. $\bm j(\bm r) = \frac{2\bm r\,(\bm d\cdot\bm r)}{r^4}- \frac{\bm d}{r^2}$ for $r>R$. Here the effective ``dipole'' moment is defined $\bm d = \alpha R^2 \frac{1}{2}(\hat{\bm z}\times \bm S)$. We also note that if the coefficients $\alpha$ and $\beta$ are large, vortex solutions for the superconducting phase $\theta$ are favored by the free energy expression~Eq.~(\ref{GL}). 

%%%%%%%%%%%%%%%%%%%%%%%%%%%%%%%%%%%%%%%%%%%%%%%%%%%%%%%%%%%%%%%%%%%%%%%%%%%
\paragraph*{Microscopic calculation.-} \label{sec:point}
%%%%%%%%%%%%%%%%%%%%%%%%%%%%%%%%%%%%%%%%%%%%%%%%%%%%%%%%%%%%%%%%%%%%%%%%%%%

%%%%%%%%%%%%%%%%%%%%%%%%%%%%%%%%%%%%%%%%%%%%%%%%%%%%%%%%%%%%%%%%%%%%%%%%%%%
\begin{figure*} \centering
(a) \includegraphics[height=0.2\linewidth]{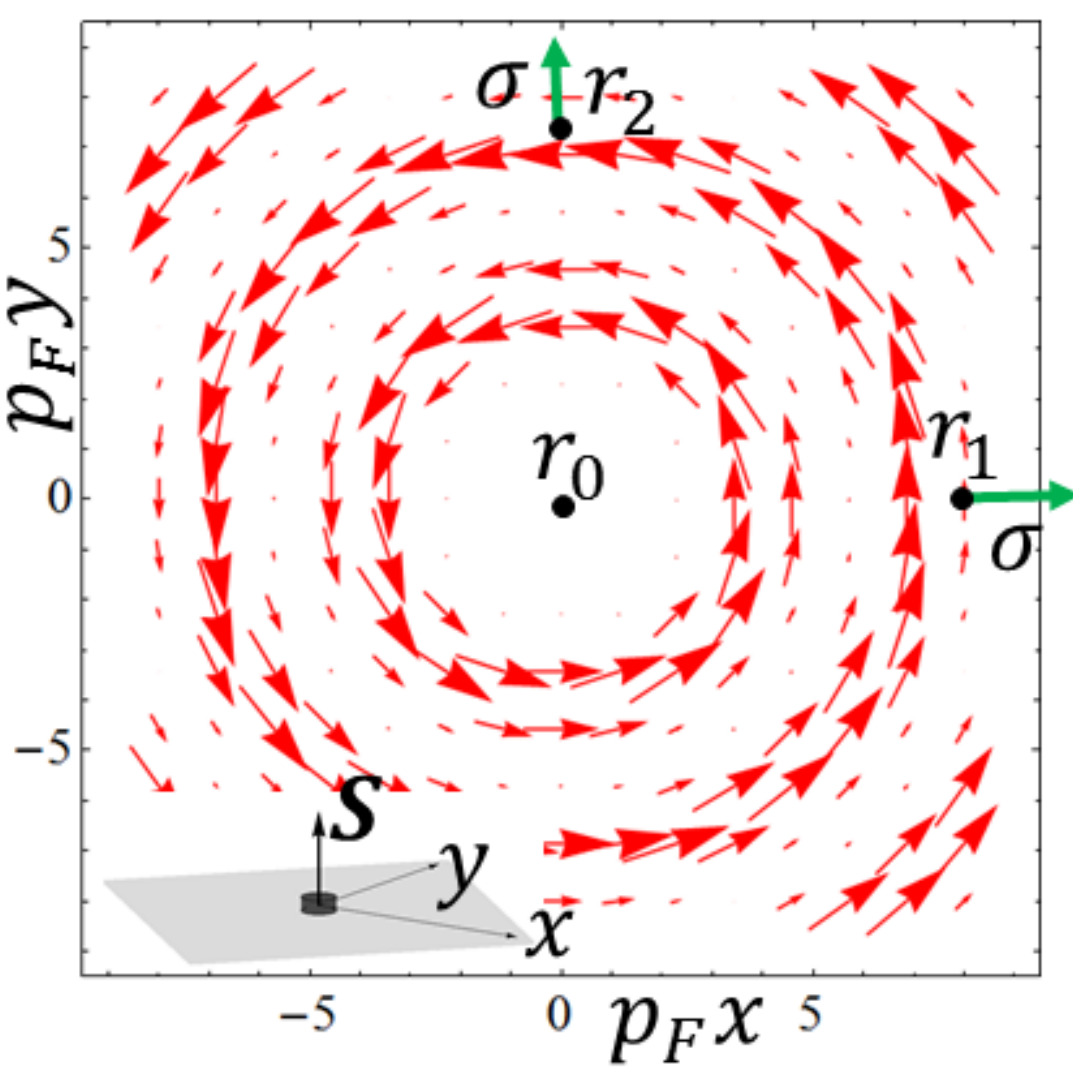} (b)  \includegraphics[height=0.2\linewidth]{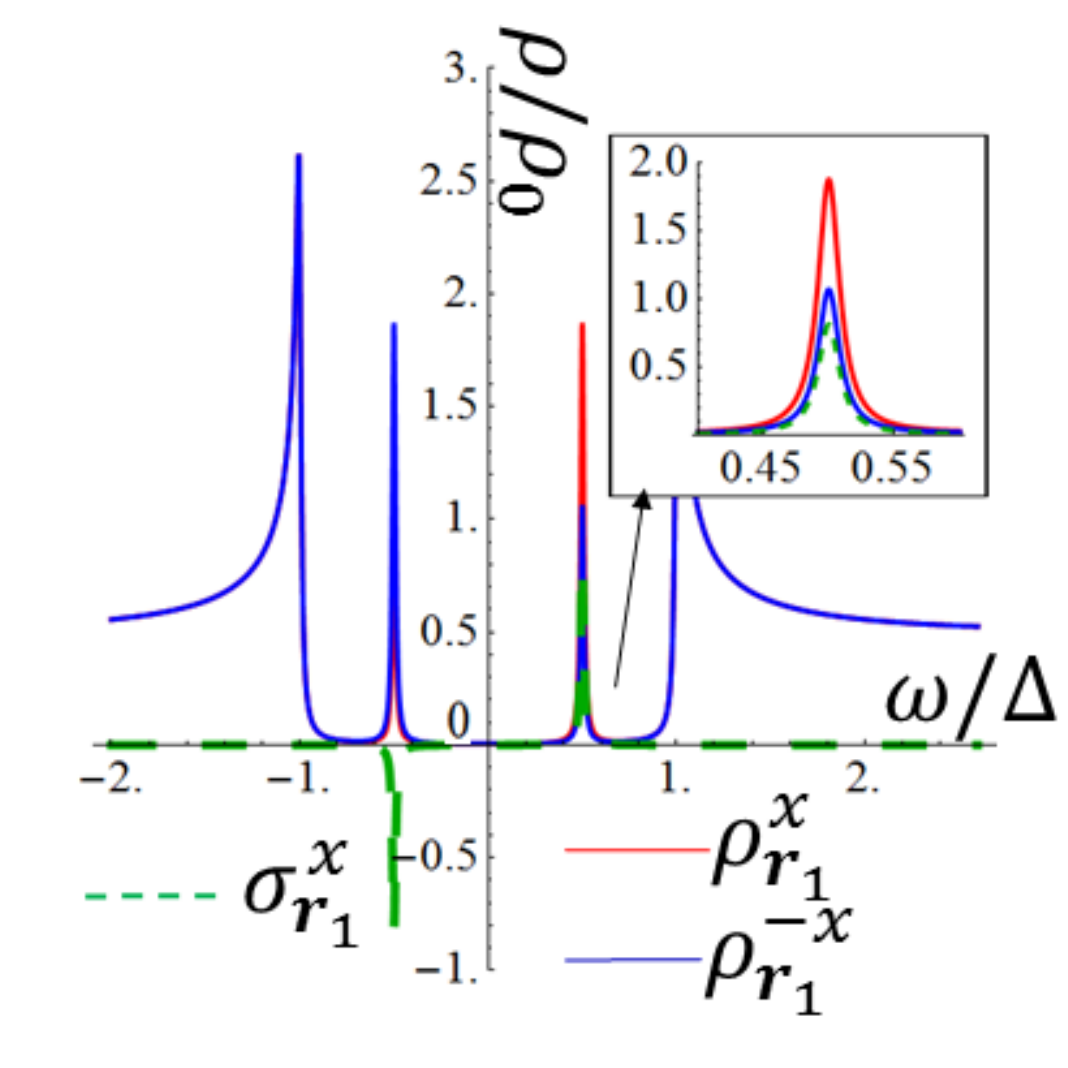} (c) \includegraphics[height=0.2\linewidth]{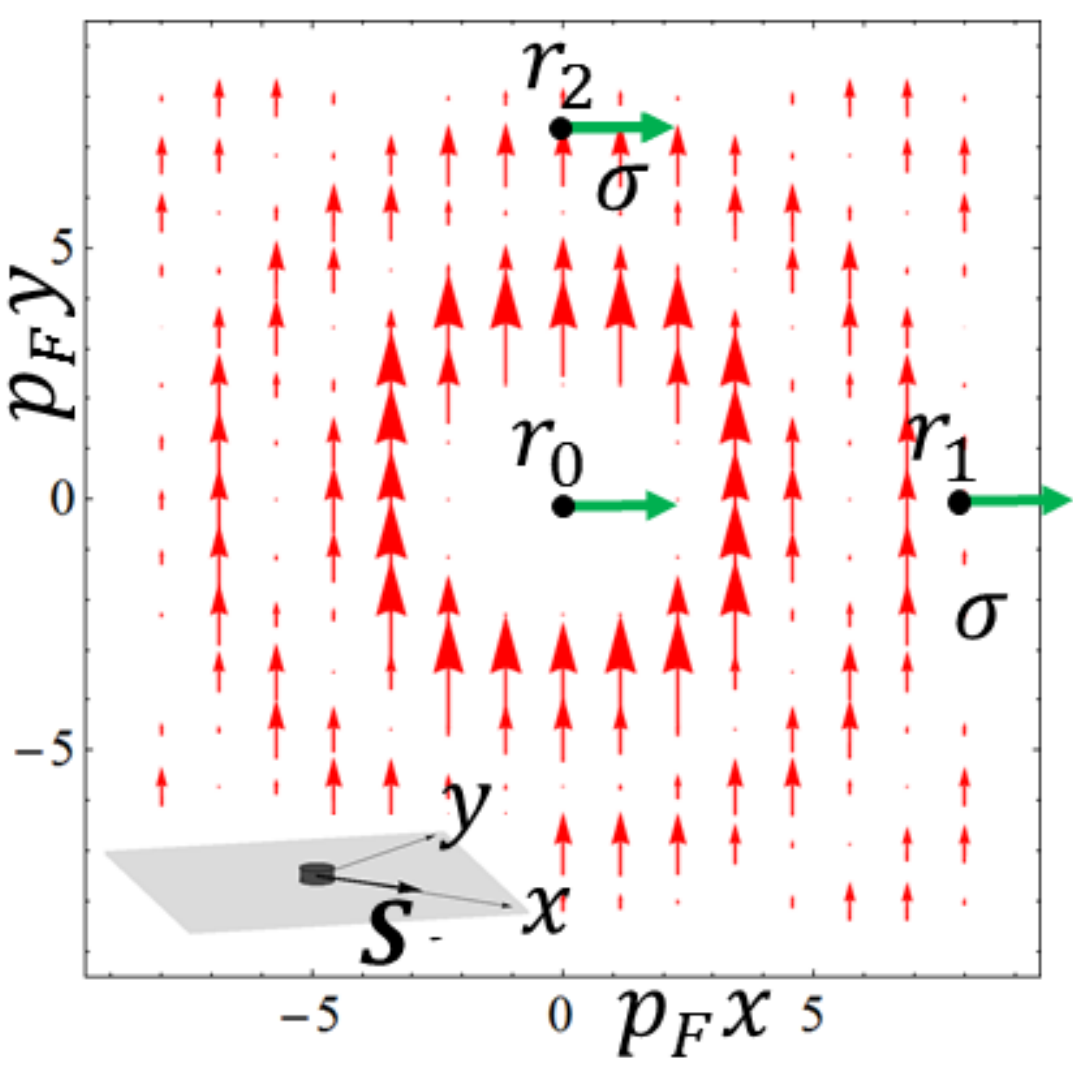} (d) \includegraphics[height=0.2\linewidth]{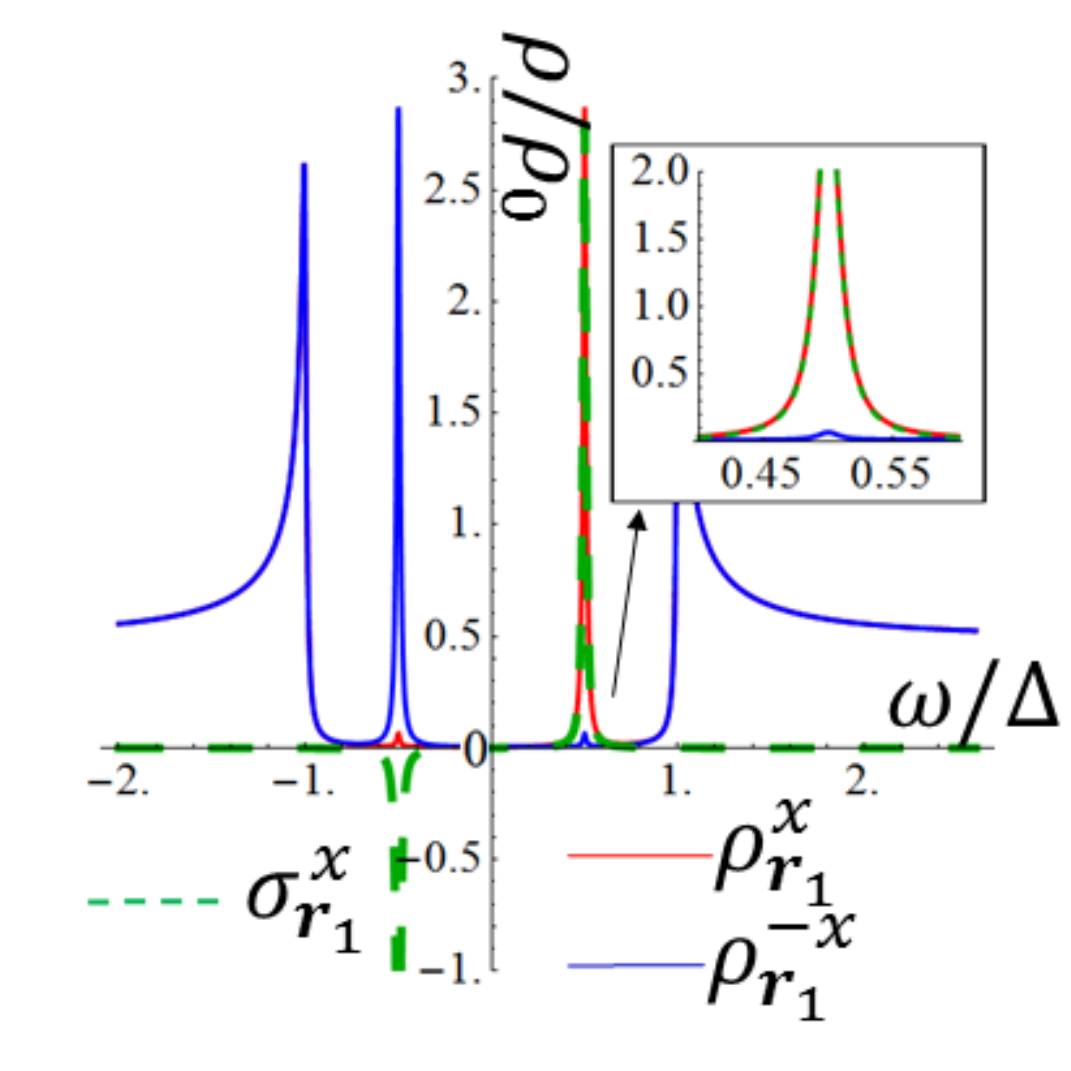}
\caption{(color online) (a),(c): Currents around a point magnetic impurity calculated using the T-matrix approach. The impurity is located at $\bm r_0=0$, and the direction of the magnetic moment $\bm S$ is shown in the inset. The vectors in green indicate the direction of the in-plane spin polarization determined from SP LDOS. The current and the spin polarization are orthogonal, which is consistent with Eq.~(\ref{main}). In order to enhance the figures, the currents are plotted away from the impurity, i.e. for $p_Fr>2.5$.  (b),(d): SP LDOS calculated at $\bm r_1$. The insets show the SP-LDOS in the vicinity of the positive YSR state. Red and blue lines correspond to SP-LDOS calculated for opposite directions $\hat{\bm x}$ and $-\hat{\bm x}$, whereas the dashed green line is the resulting spin polarization. Panels (a) and (b) correspond to the out-of-plane magnetic moment, whereas panels (c) and (d) correspond to the in-plane magnetic moment.} \label{fig2}
\end{figure*}
%%%%%%%%%%%%%%%%%%%%%%%%%%%%%%%%%%%%%%%%%%%%%%%%%%%%%%%%%%%%%%%%%%%%%%%%%%%

To complement the above GL analysis, we also study microscopically the currents generated around a single point magnetic impurity, i.e.~we set $\bm S = \bm S\,\delta(\bm r)$ in the Hamiltonian~Eq.~(\ref{H}). In contrast to the GL approach, the Green's function method, used below, allows to study effects to infinite order in $\bm S$ and also at distances smaller than the superconducting coherence length, i.e. for $r \ll \xi_{\rm sc}$. We evaluate the Green's function of the superconductor in the T-matrix approximation
\begin{align}
	G_{\bm r \bm r'}(\omega) & = g_{\bm r \bm r'}(\omega)+g_{\bm r 0}(\omega)\,T(\omega)\, g_{0 \bm r'}(\omega), \label{GF}\\ 
	T(\omega) & = \frac{-\bm S \cdot \bm \sigma}{1+\bm S \cdot \bm \sigma\,g_{00}(\omega)}.
\end{align}
The Green's function of a clean superconductor in real space at $\bm r$ is (for $r\ll\xi_{\rm sc}$) 
\begin{align}
	g_{\bm r 0}(\omega)  &=-\pi\frac{\omega+\Delta\tau_x}{\sqrt{\Delta^2-\omega^2}}  \left[f_0(r)+i(\bm \sigma\times \bm r)\,f_1(r)\right], \label{grf0}  
\end{align}
where $f_0(r) = \frac{1}{2}\left[\rho^+J_0(p_F^+ r)+\rho^-J_0(p_F^- r)\right]$ and $f_1(r) = \frac{1}{2r}\left[\rho^+J_1(p_F^+r) -\rho^-J_1(p_F^-r)\right]$,  $J_0$ and $J_1$ are Bessel functions, $p_F^\pm \approx p_F \mp \lambda m$ are the Fermi momenta of the spin-polarized Rashba bands, and $\rho^\pm \approx \rho_0 \frac{p_F^\pm}{2p_F}$ are the corresponding density of states ($\rho_0=m/\pi$). Equation~(\ref{grf0}) is calculated with the assumption $\mu\gg \Delta>0$. The second, spin-dependent, term in Eq.~(\ref{grf0}) is a consequence of the Rashba SOC and vanishes if $\lambda = 0$. The poles of the T-matrix give the energies of the impurity-induced YSR subgap states \cite{Yu, Shiba, Rusinov, Balatsky2006} 
\begin{equation}
	E^\pm_{\rm YSR} = \pm \Delta \left.\left[1-\left( \frac{\pi\rho S}{2} \right)^2\right]\right/\left[1+\left( \frac{\pi\rho S}{2} \right)^2\right], \label{Shiba}
\end{equation}
which are unaffected by the Rashba SOC \cite{Kim2014}. The energies of the YSR states, however, depend on the ferromagnetic vector magnitude $S$. For a critical value $S=2/\pi\rho$, the energies of the YSR states reach $E=0$, and the system undergoes a quantum phase transition as the two YSR states cross \cite{Salkola1997, Balatsky2006}. For simplicity, let us temporary fix $S =2/\sqrt{3}\pi\rho$, which corresponds to $E^\pm_{\rm YSR} = \pm \Delta/2$.

The current is equal to the expectation value of the velocity operator, which can be expressed using the Green's function as
\begin{align}
 \bm j(\bm r) = & \lim_{\substack{\bm r'\rightarrow \bm r\\\delta\rightarrow +0}} \int \frac{d\omega\, e^{i\omega\delta}}{2\pi i}  \label{curGF} \\
& \times{\rm Tr}\left[\frac{1+\tau_z}{2}\left( i\frac{\bm \nabla'-\bm \nabla}{2m} +\lambda\, \hat{\bm z}\times \bm\sigma\right)  G_{\bm r \bm r'}(\omega) \right]. \nonumber 
\end{align}
In addition to the usual gradient term \cite{Abrikosov} in the parenthesis, there is also a spin-dependent contribution due to the Rashba SOC. We evaluate the current~in Eq.~(\ref{curGF}) using the Green's function in Eq.~(\ref{GF}) and plot it in Figs.~\ref{fig2}(a) and (c) for the cases of out-of-plane $\bm S = S \hat {\bm z}$ and in-plane $\bm S =  S\hat {\bm x}$ moments, respectively. We note that only the pole in the T-matrix corresponding to the YSR state gives rise to nonzero currents. Both panels show concentric patterns of current centered around the impurity. In the case of the out-of-plane moment, the current circulates around the impurity. In contrast, in the case of the in-plane moment orientation, the current points predominantly in the $y$ direction. The currents shown in Fig.~\ref{fig2}(a) and (c) for point magnetic impurities agree qualitatively with the patterns obtained for the circular island within the GL theory and shown in Figs.~\ref{fig1}(b) and (c). However, in contrast with Fig.~\ref{fig1}, the currents in Fig.~\ref{fig2} display fine Friedel oscillations on the scale of $r\sim 1/p_F$. Note that the current in panel~(c) is not continuous. This can be understood by using the analogy with the Ginzburg-Landau current~(\ref{current}). For the in-plane vector $\bm S$, the current consists of the bare term $\alpha(\hat{\bm z}\times \bm S)$, as well as the condensate term $\frac{n_s}{2m}\bm\nabla\theta$. These two distinct contributions to the current are discontinuous, however, their sum is continuous. Since, the T-matrix calculation is not self-consistent, it does not take into account the reaction of the condensate that would fix the discontinuity. We discuss a fully self-consistent calculation, which demonstrates the continuity of the currents, in the next section, as well as in appendix.  

%%%%%%%%%%%%%%%%%%%%%%%%%%%%%%%%%%%%%%%%%%%%%%%%%%%%%%%%%%%%%%%%%%%%%%%%%%%
\begin{figure*} \centering
	(a) \includegraphics[height=0.2\linewidth]{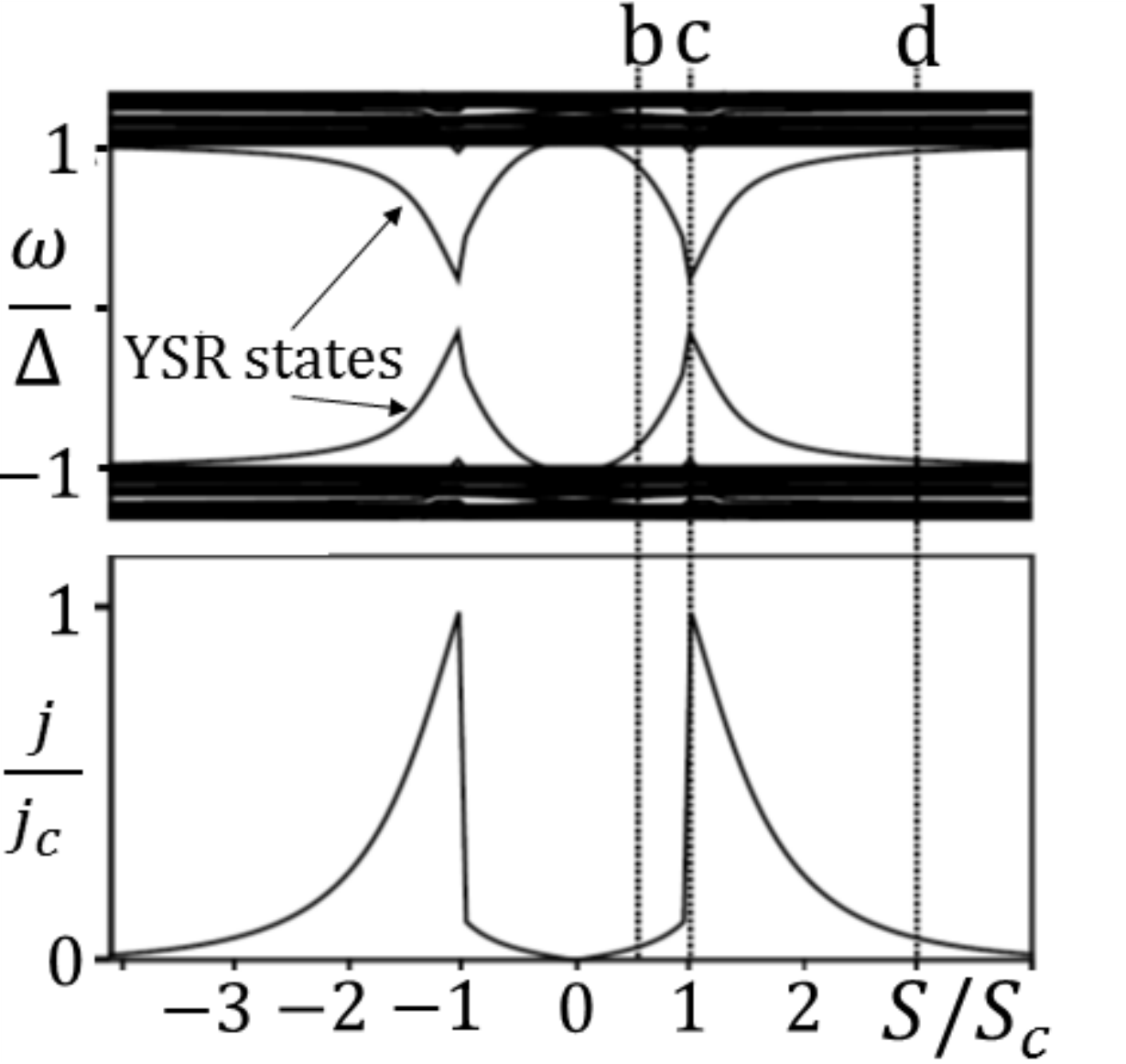} \hspace{0.1cm}(b)  \includegraphics[height=0.2\linewidth]{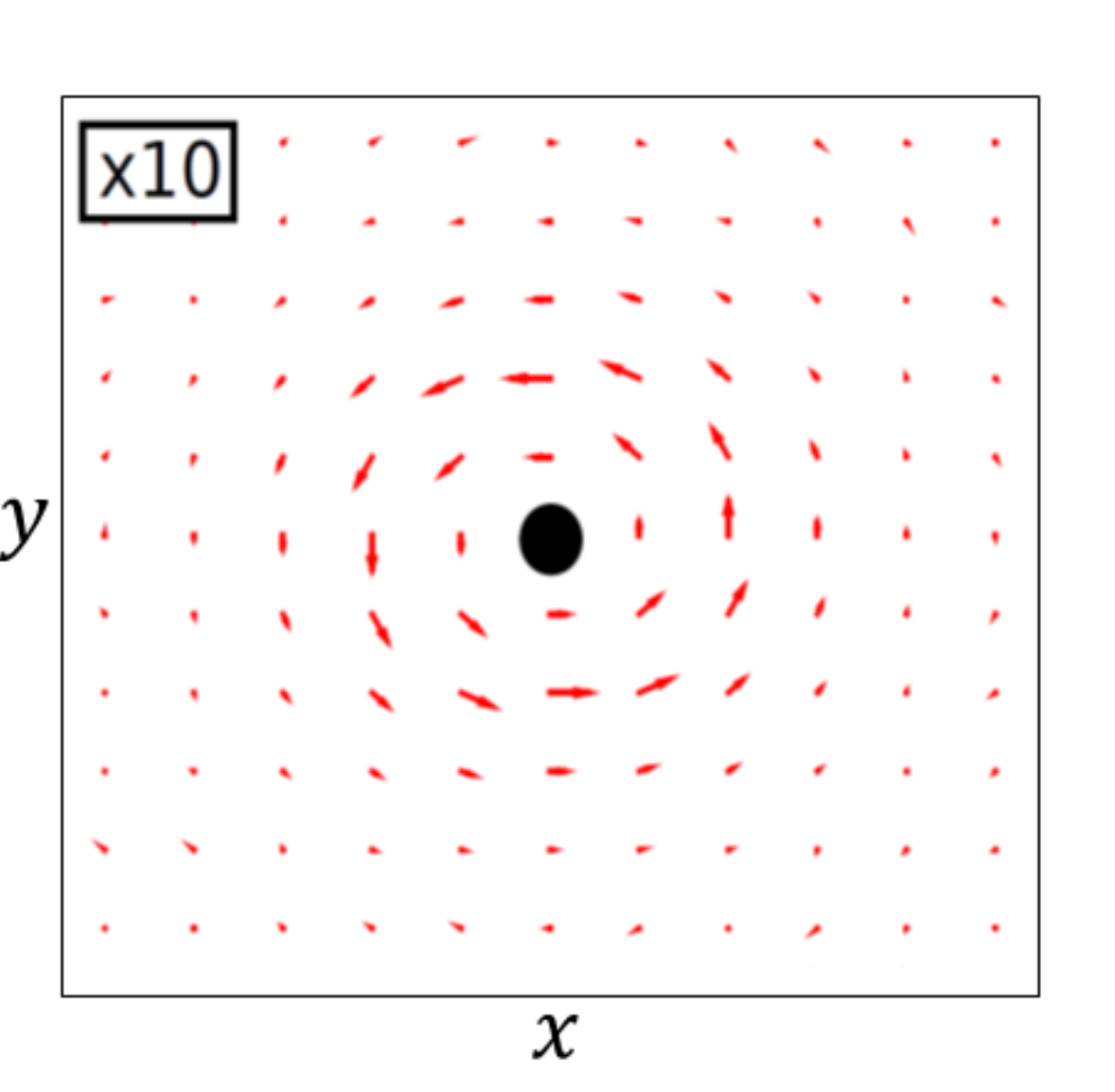}\hspace{0.1cm} (c) \includegraphics[height=0.2\linewidth]{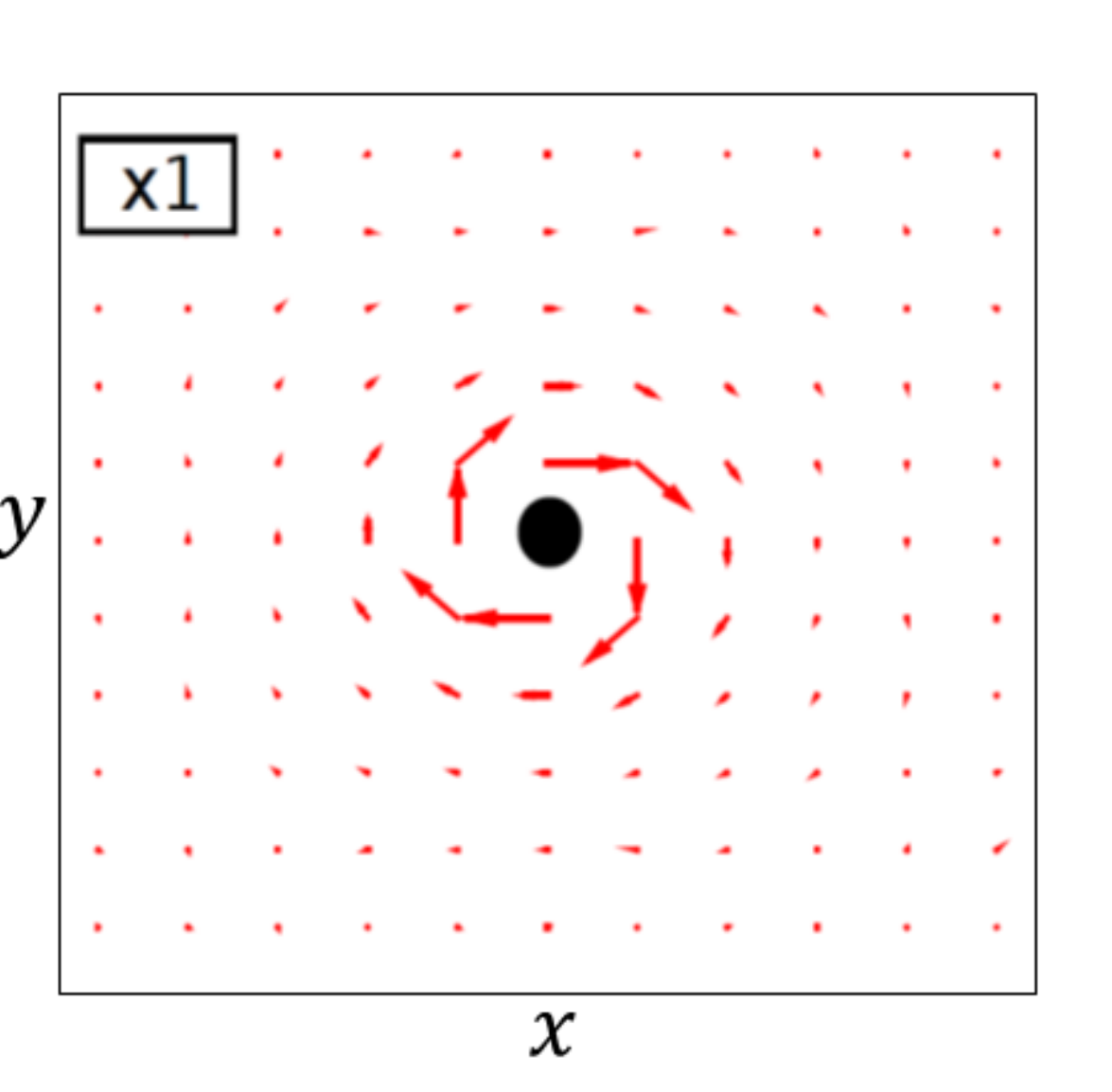}\hspace{0.1cm} (d) \includegraphics[height=0.2\linewidth]{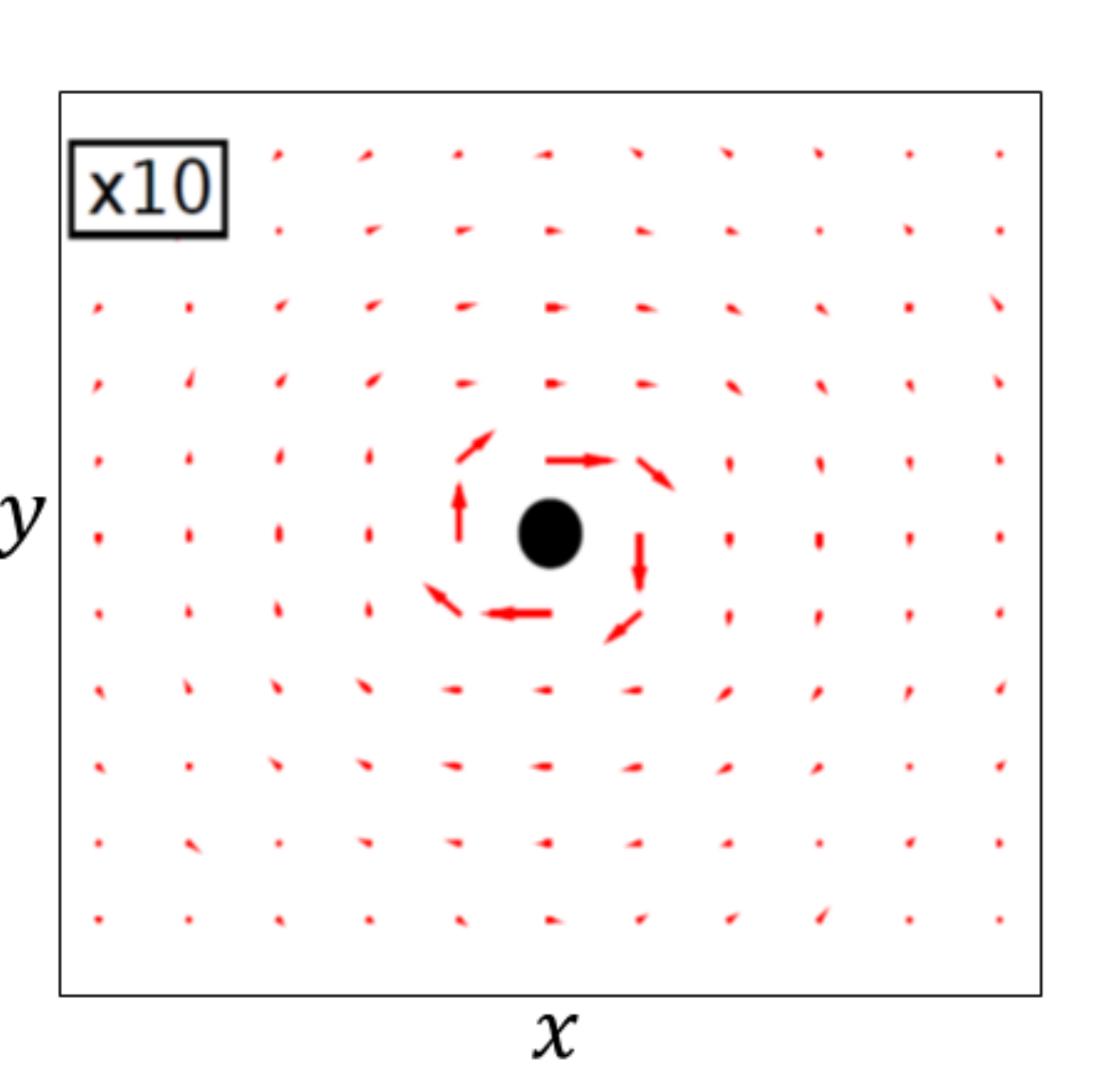} % (d)  \includegraphics[height=0.2\linewidth]{par-r1-X}% (b)  \includegraphics[width=0.35\linewidth]{par-r1-X} 
\caption{(color online) Self-consistent numerical calculation of the currents induced by a point magnetic impurity with the out-of-plane moment $\bm S = S\,\hat{\bm z}$. (a)  Bogolyubov-de Gennes spectrum (top) and magnitude of current $j$ (bottom) as a function of $S$. (b)-(d) Spatial profile of the currents plotted at discrete points on the lattice for increasing magnitude of $S$ as indicated in panel (a). The current reaches the maximum value $j_c$ and switches direction for $S=S_c$, i.e. where the YSR states cross zero energy. Note that in order to enhance visibility of the current in panels (b) and (d), the arrows representing the current are magnified tenfold as indicated by the magnification ratio in the top-left corner of each panel. } \label{fig3}
\end{figure*}
%%%%%%%%%%%%%%%%%%%%%%%%%%%%%%%%%%%%%%%%%%%%%%%%%%%%%%%%%%%%%%%%%%%%%%%%%%%

According to Eq.~(\ref{main}) the current and the spin polarization are coupled. Thus, we expect a nonzero in-plane spin polarization even away from the impurity site that sustains the nonlocal currents shown in Figs.~\ref{fig2}(a) and (c). So, we evaluate the SP LDOS using the Green's function as 
\begin{align}
	\rho^j_{\bm r}(\omega) = -\frac{1}{\pi}{\rm Im\, Tr}\,\left[\frac{1+\tau_z}{2} \frac{1+\sigma_j}{2} G_{\bm r \bm r}(\omega+i\delta) \right], \label{spldos}
\end{align}
where $j=x,y,z$ denotes the polarization axis. From the SP LDOS we also define the energy-dependent local spin polarization as  
\begin{equation}
	\sigma^j_{\bm r}(\omega)= \rho^j_{\bm r}(\omega) - \rho^{-j}_{\bm r}(\omega).
	\label{polar}
\end{equation}
In Figs.~\ref{fig2} (b) and (d), we plot both the SP LDOS and the spin polarization at the point $\bm r_1$ with solid and dashed lines, respectively.  First, consider the out-of-plane magnetic moment $\bm S =S\,\hat{\bm z}$ in panel (b). The SP LDOS peaks at the superconducting coherence peak, i.e. at $\omega = \Delta$, as well as at the subgap YSR state energy, i.e.~at $\omega = E_{\rm YSR} =  \Delta/2$. The SP LDOS corresponding to the opposite directions $j=\pm x$, shown with red and blue lines, are notably different at the YSR state. Therefore, the YSR state has a finite spin polarization along the $x$ axis, shown with a dashed green line. This feature is a consequence of the spin structure of the Green's function~Eq.~(\ref{grf0}) and vanishes in the absence of the Rashba SOC. Now, consider panel (d) corresponding to an in-plane moment $\bm S =S\,\hat{\bm x}$. The YSR state in this case has a dominating spin polarization in the $+x$ direction with only a small admixture of the opposite spin. In Figs.~\ref{fig2}(a) and (c), we plot the direction of the in-plane spin polarization for the positive YSR state $\bm \sigma_{\bm r}(E^+_{\rm YSR})= [\,\sigma^x_{\bm r}(E^+_{\rm YSR}) \,\, \sigma^y_{\bm r}(E^+_{\rm YSR})\,]$ at the point $\bm r_0=0$ as well as $\bm r_1 =7\hat{\bm x}/p_F$ and $\bm r_2 = 7\hat{\bm y}/p_F$. Note that the spin polarization of the negative YSR state is opposite, i.e. $\bm \sigma_{\bm r}(E^-_{\rm YSR})=- \bm \sigma_{\bm r}(E^+_{\rm YSR})$. The current and spin polarization are consistently orthogonal, which agrees with Eq.~(\ref{main}). So, it is possible to map the current generated by magnetic impurities and ferromagnetic islands using spin-polarized scanning tunneling microscopy (SP STM). 

%%%%%%%%%%%%%%%%%%%%%%%%%%%%%%%%%%%%%%%%%%%%%%%%%%%%%%%%%%%%%%%%%%%%%%%%%%%
\paragraph*{Self-consistent numerical modeling.-} \label{sec:Numerics}
%%%%%%%%%%%%%%%%%%%%%%%%%%%%%%%%%%%%%%%%%%%%%%%%%%%%%%%%%%%%%%%%%%%%%%%%%%%

The T-matrix approximation discussed above predicts currents which are qualitatively consistent with the GL results. However, the T-matrix approach does not capture the influence of the magnetic impurity on the superconducting order parameter. It is known that the superconducting order is strongly renormalized and may even change sign \cite{Salkola1997,Balatsky2006, Meng2015} in the vicinity of the magnetic impurity. In order to take this into account, we also perform a fully self-consistent numerical simulation\footnote{The numerical simulation is done on a square lattice with nearest-neighbor hopping $t = 1$, spin-orbit coupling  $\lambda = 0.56t$, chemical potential $\mu = -4t$. The superconducting gap is determined self-consistently using a pair potential  $v_{sc} = 5.36t$. Panels (b), (c) and (d) correspond to a magnetic impurity with $S=1.6t, 2.72t$, and $7.36t$, respectively.  For numerical reasons limiting the lattice size, the pair potential is chosen such that the superconductor coherence length is of the order of the lattice constant.} of the point magnetic impurity on a lattice \cite{Black-Schaffer2008,Kristofer2013,Kristofer2015} and show the results for an out-of-plane magnetic moment $\bm S = S \hat{\bm z}$ in Fig.~\ref{fig3}. Panels (b)-(d) show the current for increasing values of the ferromagnetic moment $S$.  Note that the Friedel oscillations are  not fully visible here since the calculation is done for a coherence length such that $\xi_{\rm sc}<1/p_F$.  In panel (a) we show the Bogolyubov-de Gennes spectrum (top) and the magnitude of the current (bottom) as a function of $S$. For small $S$ (b), the current circles around the impurity, which is consistent with both previous Figs.~\ref{fig1} and \ref{fig2}. With further increase of $S$, the current grows and ultimately undergoes a first-order discontinuous transition at a critical value of magnetic vector $S=S_c$. There, the current abruptly reverses direction, as shown in Fig.~\ref{fig3}(c) and (d), and reaches its maximal magnitude.  This is accompanied by the YSR states crossing at zero energy, and the superconducting order parameter reversing sign at the impurity site. We note that the YSR states also have a first-order avoided crossing at zero energy \cite{Salkola1997} as shown in Fig.~\ref{fig3}(a).  With further increase of $S$, superconductivity is suppressed and the currents diminish in the vicinity of the impurity. More details on the numerical simulation can be found in the appendix.

\paragraph*{Concluding remarks.-}
We have shown that superconducting currents are generated by ferromagnetic islands and single magnetic impurities in 2D superconductors with spin-orbit coupling.  The currents originate from the magnetoelectric effect and are a direct consequence of combining SOC and magnetism. The discussed currents are unavoidable in ferromagnet-superconductor heterostructures, which have been proposed as a platform for topological superconductivity with the Majorana boundary states \cite{Lutchyn2010,Oreg2010, Sau2010, Choy2011, Nadj-Perge2013, Klinovaja2013, Vazifeh2013, Braunecker2013, Pientka2013, Nakosai2013, Poyhonen2014, Kim2014a,  Reis2014, Yazdani2014, Brydon2015, Rontynen2014, Li2015}. We find a strong dependence of both the spatial pattern and magnitude of the currents on the direction of the ferromagnetic moment. The currents are localized on the scale of the coherence length in the case of the out-of-plane local magnetic moment, whereas the currents have a dipolar power law decay in the case of the in-plane magnetic moment. The presence of these non-local currents may induce long-range interactions between local magnetic moments on a superconductor \cite{Yao2014}, which could qualitatively change the behavior of the Majorana modes in such systems. Furthermore, by analyzing the currents in detail, we find that they are carried by the subgap YSR states induced by point magnetic impurities. The YSR states are spin-polarized, and the current is orthogonal to the local spin polarization. Moreover, the current magnitude peaks sharply at the phase transition, where the YSR states cross at zero energy. Thus, by using SP STM it should be possible to map out the currents as well as detect the phase transition, which is paramount for finding TS and the Majorana modes.

%{\it Acknowledgment}. 
We thank G.~Volovik, M.~Eschrig, Y.~Kedem, and C.~Triola for useful discussions. This work was supported by the European Research Council (ERC) DM-321031 and the US DOE BES E304 (S.S.P. and A.V.B.) and the Swedish Research Council (Vetenskapsr\aa det), the G\"oran Gustafsson Foundation, and the Swedish Foundation for Strategic Research (SSF) (K.B. and A.B.-S.). 

\newpage
%%%%%%%%%%%%%%%%%%%%%%%%%%%%%%%%%%%%%%%%%%%%%%%%%%%%%%%%%%%%%%%%%%%%%%%%%%%%%
%\bibliographystyle{apsrev4-1}
\bibliography{ImpurityCurrents}
%%%%%%%%%%%%%%%%%%%%%%%%%%%%%%%%%%%%%%%%%%%%%%%%%%%%%%%%%%%%%%%%%%%%%%%%%%%%%

\appendix

%%%%%%%%%%%%%%%%%%%%%%%%%%%%%%%%%%%%%%%%%%%%%%%%%%%%%%%%%%%%%%%%%%%%%%%%%%%%
\section{Derivation of the extra terms in the Ginzburg-Landau theory} \label{appendix:Derivation}
%%%%%%%%%%%%%%%%%%%%%%%%%%%%%%%%%%%%%%%%%%%%%%%%%%%%%%%%%%%%%%%%%%%%%%%%%%%%
The non-local coupling of the vector $\bm {\mathcal A}= (\mathcal A_x,\mathcal A_y)= \bm A+\frac{\bm\nabla\theta}{2}$ and ferromagnetic polarization $\bm S$ in  2D momentum $\bm q=(q_x,q_y)$ space is given by the extra term in the free energy
\begin{equation}
	F_{\rm extra} = -\int\limits_{\bm q} S_a(\bm q)K_{ab}(\bm q)\mathcal A_b(-\bm q). 
	\label{extra}
\end{equation}
The tensor $K_{ab}$ is obtained by integrating out the fermions
\begin{equation}
	K_{ab}(\bm q) = \frac{1}{2} \int\limits_{\omega,\bm p}  {\rm Tr} \left[ \sigma_a\,g_{\bm p+\frac{\bm q}{2}}\left(i\omega\right)\,v_b\, g_{\bm p-\frac{\bm q}{2}}\left(i\omega \right)\right] \label{Kab}
\end{equation}
where the integration variables are given below the integral sign for brevity, i.e. $\int\limits_{\omega,\bm p}=\int \frac{d\omega d^2p}{\left(2\pi\right)^3}$. The integration over continuous frequency $\omega$ corresponds to zero temperature $T=0$. The factor $1/2$ takes care of the doubling of degrees of freedom in the 4-by-4 Bogolyubov-de Gennes representation. The Green's function as well as the velocity operators are defined as
\begin{align}
	& g_{\bm p}(i\omega) = \frac{1}{i\omega-[\xi(\bm p)-\lambda\left( \bm\sigma\times\bm p\right)_z]\tau_z-\Delta\tau_x}, \label{gf}\\
	& \bm v =  \frac{\bm p}{m}+\lambda(\hat{\bm z}\times\bm \sigma). \label{vel}
\end{align}
We expand the tensor $K_{ab}(q)$ up to the first order in $q$
\begin{equation}
	K_{ab}(\bm q) = K_{ab}(0)+q^c\,\partial_{q_c}K_{ab}(0)+\mathcal O(q^2). \label{exp}
\end{equation}
and find the tensors $K_{ab}(0)$ and $\partial_cK_{ab}(0)$ in the following two subsections
\begin{align}
	K_{ab}(0) &=  \epsilon_{zba}\,\alpha,\quad \alpha = \frac{\lambda m}{2\pi}, \label{k0} \\ 
	\partial_cK_{ab}(0) & =  i\delta_{az}\epsilon_{zcb}\,\beta,\quad \beta = \frac{m^2 \lambda^2}{4\pi p_F^2}. \label{k1}
\end{align}
Here, the coefficients $\alpha$ and $\beta$ were evaluated in the limit of the small superconducting gap $\Delta$ and large Fermi momentum $p_F=\sqrt{2m\mu}$, i.e. $\lambda p_F\gg m\lambda^2\gg \Delta$. It may also be useful to instead express the coefficients using the density of states $\rho_0 = m/\pi$ and Rashba momentum ${p_R = m\lambda}$ as
\begin{equation}
	\alpha =  \rho_0\lambda\, \frac{1}{2}, \quad \beta = \rho_0 \lambda\,\frac{p_R}{4p_F^2}.
\end{equation}
We can now substitute Eqs.~(\ref{k0}) and (\ref{k1}) in Eq.~(\ref{exp}) and rewrite Eq.~(\ref{extra}) in real space $\bm r = (x,y)$ as
\begin{align}
	F_{\rm extra} &= \int\limits_{\bm r}  \alpha\, (\hat{\bm z}\times\bm S) \cdot \bm{\mathcal A}+\beta\, (\bm \nabla S_z \times  \hat{\bm z}) \cdot \bm{\mathcal A}. 
	\label{extraR}
\end{align}
The term $\alpha$ was also recently derived in Ref.~\cite{Malshukov2014}. We are not aware of a previous derivation of the term $\beta$.

%%%%%%%%%%%%%%%%%%%%%%%%%%%%%%%%%%%%%%%%%%%%%%%%%%%%%%%%%%%%%%%%%%%%%%%%%%%%
\subsection{Derivation of the tensor $ K_{ab}(0)$.} \label{appendix:alpha}
%%%%%%%%%%%%%%%%%%%%%%%%%%%%%%%%%%%%%%%%%%%%%%%%%%%%%%%%%%%%%%%%%%%%%%%%%%%%

Let us  calculate the first term in this expansion, i.e. 
\begin{equation}
	K_{ab}(\bm 0) = \frac{1}{2} \int\limits_{\omega,\bm p}  {\rm Tr} \left[ \sigma_a\,g_{\bm p}\left(\bm p\right)\,v_b\, g_{\bm p}\left(\bm p\right)\right]. \label{Kab0}
\end{equation}
We define the operators 
\begin{equation}
	\Pi_{\pm} = \frac{1}{2} \left[1 \pm \left( \bm\sigma\times\hat{\bm p}\right)_z \right],\,\,\hat{\bm p} = \frac{\bm p}{p},
	\label{projector}
\end{equation}
which project onto the spin eigenstates of the Rashba coupling $\lambda\left(\bm\sigma\times\bm p\right)_z$ corresponding to the two eigenstates $\pm \lambda p$. We expand the identity operator $\bm I$ via the projection operators~(\ref{projector}) as $\bm I = \Pi_++\Pi_-$ and simplify the Green's function 
\begin{equation}
	g_{\bm p}(i\omega) = \Pi_+\frac{i\omega+\xi_+(\bm p)\tau_3+\Delta\tau_1}{(i\omega)^2-E_+^2(\bm p)}+\Pi_-\frac{i\omega+\xi_-(\bm p)\tau_3+\Delta\tau_1}{(i\omega)^2-E_-^2(\bm p)}, \label{gExp}
\end{equation}
where we defined the energies due to the Rashba splitting 
\begin{equation}
	\xi_{\pm}(\bm p) = \xi (\bm p)\pm\lambda p,\,\, E_\pm(\bm p) = \sqrt{\xi^2_\pm (\bm p)+\Delta^2}.
\end{equation}
We substitute Eq.~(\ref{gExp}) in Eq.~(\ref{Kab0}) and obtain four terms. Only the terms containing both $\Pi_+$ and $\Pi_-$ have distinct poles, which produce a non-vanishing result upon the frequency integration. Thus we obtain\begin{align}
	& K_{ab}( 0) = \frac{1}{2}\int\limits_{\bm p} \,{\rm Tr}_\sigma\left\{\sigma_a\Pi_+ v_b \Pi_-+\sigma_a\Pi_- v_b \Pi_+\right\} \nonumber \\
	& \times \int\limits_\omega \frac{{\rm Tr}_\tau\{[i\omega+\xi_+(\bm p)\tau_3+\Delta\tau_1][i\omega+\xi_-(\bm p)\tau_3+\Delta\tau_1]\}}{[(i\omega)^2-E_+^2(\bm p)][(i\omega)^2-E_-^2(\bm p)]}.  \label{Kab2}
\end{align}
where the first trace is over the spin space, whereas second trace is over the Nambu space.  The trace in the second line of Eq.~(\ref{Kab2}) gives $2[-\omega^2+\xi^+(\bm p)\xi^-(\bm p)+\Delta^2]$. Using the angular integration in momentum space, we simplify the trace over spin matrices in the first line of Eq.~(\ref{Kab2}) to $\lambda\epsilon_{zab}$. Then, we integrate over frequency and obtain
\begin{align}
	K_{ab}(0) &= \epsilon_{zba}\, \frac{\lambda}{2}\int\limits_{\bm p} \frac{E_+(\bm p)E_-(\bm p)-\xi_+(\bm p)\xi_-(\bm p)-\Delta^2}{E_+(\bm p)E_-(\bm p)[E_+(\bm p)+E_-(\bm p)]}.\label{coef0}        
\end{align}
We can evaluate the integral in various limiting cases
\begin{align}
	K_{ab}(0) =  \epsilon_{zba}\, \frac{m\lambda}{2\pi} \left\{\begin{array}{ll}
  				 2\lambda^2 p_F^2/3\Delta^2, & \mu\gg\Delta\gg \lambda p_F\gg m\lambda^2, \\
		           	 1, 				& \lambda p_F\gg m\lambda^2\gg \Delta,
			 \end{array}\right.  \label{K0f}
\end{align}
where $p_F = \sqrt{2m\mu}$ is the Fermi momentum. The first line of Eq.~(\ref{K0f}) can also be obtained from Eq.~(7) of Ref.~\cite{Malshukov2014}.

% UNCOMMENT BELOW TO SEE EPLANATION ABOUT THE INTEGRAL IN MOMENTUM space

%Let us evaluate Eq.~(\ref{coef0}) in various limiting cases. First, consider the case of small spin-orbit coupling, i.e. $\Delta\gg \lambda p_F> m\lambda^2$, where $p_F = \sqrt{2m\mu}$. The integrand is peaked around $p=p_F$, so, we linearize $\xi=v_F (p-p_F)$ and evaluate the integral 
%\begin{align}
%	K_{ab}(0) &=  \epsilon_{zab}\,\lambda\rho_0\frac{2\lambda^2 p_F^2}{3\Delta^2}, \label{coef00}
%\end{align}
%where $\rho_0 = m/\pi$ is the density of states. The same result is obtained after Matsubara summation is made in Eq.~(7) of Ref.~\cite{Malshukov2014}. In the opposite limit, where the superconducting gap is small $\lambda p_F> m\lambda^2\gg \Delta$, we drop $\Delta$ in Eq.~(\ref{coef0}), and the integrand simplifies to $\theta(-\xi_-(\bm p)\xi_+(\bm p))/p\lambda$. The Heaviside theta function is non-zero for momenta $p_F^+<p<p_F^-$, which satisfy $\xi_\pm(p_F^\pm)=0$, and the integral can be easily evaluated 
%\begin{align}
%	K_{ab}(0) &=  \epsilon_{zab}\,\frac{p_F^--p_F^+}{2\pi} =  \epsilon_{zab}\,\lambda\rho_0. \label{coef000}
%\end{align}

%%%%%%%%%%%%%%%%%%%%%%%%%%%%%%%%%%%%%%%%%%%%%%%%%%%%%%%%%%%%%%%%%%%%%%%%%%%%
\subsection{Derivation of the tensor $ \partial_{q_c}K_{ab}(0)$.} \label{appendix:beta}
%%%%%%%%%%%%%%%%%%%%%%%%%%%%%%%%%%%%%%%%%%%%%%%%%%%%%%%%%%%%%%%%%%%%%%%%%%%%

We expand Eq.~(\ref{Kab}) and obtain the first order coefficient 
\begin{align*}
	\partial_{q_c}K_{ab} = \frac{1}{4} \int\limits_{\omega,\bm p} {\rm Tr}\, \left\{ \sigma_a\left[\partial_{p_c} g_{\bm p}(i\omega)\,v_b\, g_{\bm p}(i\omega)\right.\right.\\
	\left.\left.- g_{\bm p}(i\omega)\,v_b\,\partial_{p_c} g_{\bm p}(i\omega)\right]\right\}.
\end{align*}
We integrate by parts to shift the position of the derivate $\partial_{p_c}$, use the identity $\partial_{p_c}g=g\,\tau_z\, v_c\,g$ (which follows from Eqs.~(\ref{gf}) and (\ref{vel})) and obtain 
\begin{align*}
	\partial_{q_c}K_{ab} &= \frac{1}{2} \int\limits_{\omega,\bm p} {\rm Tr}\, \left\{ \sigma_a\,\partial_{p_c}  g_{\bm p}(i\omega)\,v_b\, g_{\bm p}(i\omega)\right\} \\
			     &= \frac{1}{2}  \int\limits_{\omega,\bm p} {\rm Tr}\, \left\{ \sigma_a\, g_{\bm p}(i\omega)\, \tau_z\, v_c \, g_{\bm p}(i\omega)\,v_b\, g_{\bm p}(i\omega)\right\}
\end{align*}
Below, we omit additional terms containing powers of the Green's function $g^n$ under the trace because they vanish upon the frequency integration (since poles lie on the same side of the imaginary plane). We substitute the expressions for the velocity Eq.~(\ref{vel}), omit the terms that vanish under the angular integration in the momentum space and simplify the equation to
\begin{align}
&	\partial_{q_c}K_{ab}(0) = \nonumber \\
&	\frac{\lambda^2 \epsilon_{z\tilde cc}\epsilon_{z\tilde bb}}{2}  \int\limits_{\omega,\bm p} {\rm Tr}\, \left\{ \sigma_a\, g_{\bm p}(i\omega)\, \tau_z\, \sigma_{\tilde c} \, g_{\bm p}(i\omega)\,\sigma_{\tilde b}\, g_{\bm p}(i\omega)\right\} \label{interm}
\end{align}
Here, we dropped the term containing $p_bp_c{\rm Tr}\{\sigma_a g_{\bm p}(i\omega)\tau_z g^2_{\bm p}(i\omega)\}$ which vanishes upon the frequency integration. We substitute the expansion of the Green's functions Eq.~(\ref{gExp}) in Eq.~(\ref{interm}) and traces over the Nambu and spin matrices decouple such that
\begin{align*}
	\sum\limits_{s_1,s_2,s_3 = \pm} & \frac{{\rm Tr}_\sigma\, \left\{ \sigma_a\,\Pi_{s_1}\, \sigma_{\tilde c}\, \Pi_{s_2}\,\sigma_{\tilde b}\, \Pi_{s_3}\right\}}{[(i\omega)^2-E_{s_1}^2(\bm p)][(i\omega)^2-E_{s_2}^2(\bm p)][(i\omega)^2-E_{s_3}^2(\bm p)]}\\
	\times{\rm Tr}_\tau\, & \left\{ [i\omega+\xi_{s_1}(\bm p)\tau_z+\Delta\tau_x]  \tau_z [i\omega+\xi_{s_2}(\bm p)\tau_z+\Delta\tau_x] \right. \\ 
&       \times\left.[i\omega+\xi_{s_3}(\bm p)\tau_z+\Delta\tau_x]\right\}.
\end{align*}
First we evaluate the trace over the spin matrices 
\begin{align*}
	{\rm Tr}_\sigma\, \left\{ \sigma_a\,\Pi_{s_1}\, \sigma_{\tilde c}\, \Pi_{s_2}\,\sigma_{\tilde b}\, \Pi_{s_3}\right\}\rightarrow(1-s_1 s_3)\frac{i\,\epsilon_{a\tilde c\tilde b}}{4},
\end{align*}
where the angular integration in the momentum space was invoked to simplify the expression. The remaining trace over the Nambu matrices can also be evaluated as
\begin{align*}
	&{\rm Tr}_\tau\,  \left\{ [i\omega+\xi_{s_1}(\bm p)\tau_z+\Delta\tau_x]  \tau_z [i\omega+\xi_{s_2}(\bm p)\tau_z+\Delta\tau_x] \right. \\ 
&       \times\left.[i\omega+\xi_{s_3}(\bm p)\tau_z+\Delta\tau_x]\right\} = -2\omega^2\left[ \xi_{s_1}(\bm p)+\xi_{s_2}(\bm p)+\xi_{s_3}(\bm p)\right] \\
&+ 2\xi_{s_1}(\bm p)\xi_{s_2}(\bm p)\xi_{s_3}(\bm p)+2\Delta^2\left[ \xi_{s_1}(\bm p)+\xi_{s_2}(\bm p)-\xi_{s_3}(\bm p)\right].
\end{align*}
We integrate over frequency, substitute all terms in Eq.~(\ref{interm}) and obtain 
\begin{align}
	&	\partial_{q_c}K_{ab}(0) = \label{coef1}\\
&	\frac{i\delta_{az}\epsilon_{zcb}}{8}\int \limits_{\bm p} \frac{\lambda \Delta^2}{\xi^2(\bm p) p} \left[ \frac{\xi(\bm p)\lambda p}{E_+^3(\bm p)}+\frac{\xi(\bm p)\lambda p}{E_-^3(\bm p)}+\frac{1}{E_+(\bm p)}-\frac{1}{E_-(\bm p)} \right]\nonumber
\end{align}
Similar to Eq.~(\ref{K0f}), we evaluate the integral in the limit of small $\Delta$ and obtain
\begin{align}
		\partial_{q_c}K_{ab}(0) = i\delta_{az}\epsilon_{zcb}\, \frac{m^2 \lambda^2}{4\pi p_F^2},\,\,\, \lambda p_F\gg m\lambda^2\gg \Delta. \label{K1f}
\end{align}
% UNCOMMENT BELOW TO SEE EPLANATION ABOUT THE INTEGRAL IN MOMENTUM space

%In the limit $\mu\gg \Delta$, we evaluate the integral and obtain 
%\begin{align}
%	&	\partial_{q_c}K_{ab}(0) = i\delta_{az}\epsilon_{zcb}\, \frac{m^2 \lambda^2}{4\pi p_F^2} \label{coef11}\\	&       \times \left[\frac{1}{1-\frac{m^2\lambda ^2}{p_F^2}}-\frac{\Delta ^2}{\Delta ^2+\lambda^2 p_F^2}+\frac{\lambda p_F  \tanh ^{-1}\left(\frac{\lambda p_F }{\sqrt{\Delta ^2+\lambda^2 p_F^2}}\right)}{\left(\Delta ^2+\lambda^2 p_F^2\right)^{3/2}}\right]\nonumber
%\end{align}

%%%%%%%%%%%%%%%%%%%%%%%%%%%%%%%%%%%%%%%%%%%%%%%%%%%%%%%%%%%%%%%%%%%%%%%%%%%%
\section{Ginzburg-Landau solution of currents induced by a ferromagnetic disc}
%%%%%%%%%%%%%%%%%%%%%%%%%%%%%%%%%%%%%%%%%%%%%%%%%%%%%%%%%%%%%%%%%%%%%%%%%%%%
In this section, we provide details of the calculation of the current
\begin{equation}
	\bm j = \left.\frac{\delta F}{\delta \bm A}\right|_{\bm A = 0}= \frac{n_s}{2m}\bm\nabla\theta+\alpha(\hat{\bm z}\times \bm S )+\beta (\hat{\bm z}\times\bm\nabla S_z). \label{current1}
\end{equation}
around a ferromagnetic region with disc geometry
\begin{equation}
	\bm S(\bm r) = \bm S\,\theta_H(R-r), \label{island} 
\end{equation}
where the index in $\theta_H$ denotes the Heaviside theta function to contrast it with the phase of the condensate $\theta$. Let us first consider the case where the spin is out-of-plane, i.e. $\bm S = S \hat{\bm z}$. Then, taking into account Eq.~(\ref{island}), Eq. (\ref{current1}) becomes 
\begin{align}
\bm j(\bm r) &= \left.\frac{n_s}{2m}\bm\nabla\theta+\beta [\hat{\bm z}\times\bm\nabla S_z(\bm r)]\,\right|_{\theta = 0} \label{currentZ} \\
	&=  \beta S \,(\hat{\bm r} \times \hat{\bm z})\,\delta (r-R).  \label{currentZ1}
\end{align}
The current is thus localized around the boundary of the ferromagnetic region as shown in Fig.~\ref{fig1}(b). Note that Eq.~(\ref{currentZ1}) corresponds to vanishing superconducting phase $\theta=0$, which is valid for small constant $\beta$. For larger values of $\beta$, the vortex configuration of the superconducting phase, i.e. $\theta(\bm r) = {\rm atan}(y/x)$, minimizes the free energy and the full expression~Eq.~(\ref{currentZ}) for the current must be used.

Now let us consider the case of an in-plane ferromagnetic vector, i.e. $\bm S = S\,\hat {\bm x}$. Then, the current~Eq.~(\ref{current1}) becomes
\begin{align}
	\bm j(\bm r) &= \frac{n_s}{2m}\bm\nabla\theta+\alpha\, [\hat{\bm z}\times \bm S(\bm r) ]. \label{currentX} 
\end{align}
Notice here that the Euler-Lagrange equation for the superconducting phase $\theta$ gives the continuity equation for the current 
\begin{equation}
	0 = \bm \nabla\cdot \bm j =   \frac{n_s}{2m}\Delta\theta+\alpha\, \bm\nabla\cdot[\hat{\bm z}\times \bm S(\bm r)].
	\label{cont}
\end{equation}
We rewrite Eq.~(\ref{cont}) as $\frac{n_s}{2m\alpha}\,\Delta\theta=\hat{\bm z}\cdot[\bm\nabla\times \bm S(\bm r)] =-\hat{\bm z}\cdot(\hat {\bm r}\times \bm S)\delta(r-R)$, use the Green's function $G_L(\bm r) = \frac{1}{2\pi}\,{\rm ln}(r)$ for the 2D Laplace operator, which satisfies $\Delta G_L(\bm r) = \delta^2(\bm r)$, and find the solution ${\theta (\bm r) = -\frac{2m\alpha\epsilon_{zab} S_{0b}}{n_s} \int d^2r'\,G_L(\bm r-\bm r') \hat{r}'_a\, \delta(r'-R)}$, which after integration gives a simple result 
\begin{align}
	\theta (\bm r) % &= \frac{m\alpha}{n_s}\, {\bm r}\cdot(\bm S\times \hat{\bm z}) \, \frac{1-\sqrt{1-\left( \frac{2rR}{R^2+r^2} \right)^2}}{\frac{2r^2}{R^2+r^2}}  \\
	&=  \frac{m\alpha}{2n_sr^2}\,{\bm r}\cdot(\bm S\times \hat{\bm z})\,\left[ R^2+r^2-\left|R^2-r^2\right| \right] \label{th}\\
	&= \frac{m\alpha}{n_s}\,{\bm r}\cdot(\bm S\times \hat{\bm z})\,\left\{\begin{array}{ll}
  				 1, & r < R,  \\
				 \frac{R^2}{r^2}, & r > R.
			 \end{array}\right. \label{thl}
\end{align}
Using Eqs.~(\ref{th}) and (\ref{island}) we calculate the current~(\ref{currentX})
\begin{align}
     \bm j(\bm r) = \left\{\begin{array}{ll}
	 		        \frac{\bm d}{R^2}, & r < R,  \\
				\frac{2\bm r\,(\bm d\cdot\bm r)}{r^4}- \frac{\bm d}{r^2}, & r> R,
			 \end{array}\right. 
\end{align}
where the ``dipole'' moment is defined as ${\bm d= \alpha R^2 (\hat{\bm z}\times \bm S)/2}$.  

Note that the problem of calculating the current discussed above is formally equivalent to the magnetostatics problem of calculating the magnetic field $\bm B = \bm H+4\pi\bm M$ induced by the ferromagnet of magnetization $\bm M$. In the magnetostatics problem, the constant magnetization $\bm M$ is equivalent to $\alpha\, [\hat{\bm z}\times \bm S(\bm r)]$, and the magnetic field strength $\bm H$ is the term $\frac{n_s}{2m}\bm\nabla\theta$ in Eq.~(\ref{currentX}). The divergenceless magnetic induction $\bm B$ is equivalent to the current $\bm j$.

%%%%%%%%%%%%%%%%%%%%%%%%%%%%%%%%%%%%%%%%%%%%%%%%%%%%%%%%%%%%%%%%%%%%%%%%%%%%
\section{T-matrix calculation} \label{sec:tmatrix}
%%%%%%%%%%%%%%%%%%%%%%%%%%%%%%%%%%%%%%%%%%%%%%%%%%%%%%%%%%%%%%%%%%%%%%%%%%%%

%%%%%%%%%%%%%%%%%%%%%%%%%%%%%%%%%%%%%%%%%%%%%%%%%%%%%%%%%%%%%%%%%%%%%%%%%%%
\begin{figure} \centering
(a) \includegraphics[width=0.7\linewidth]{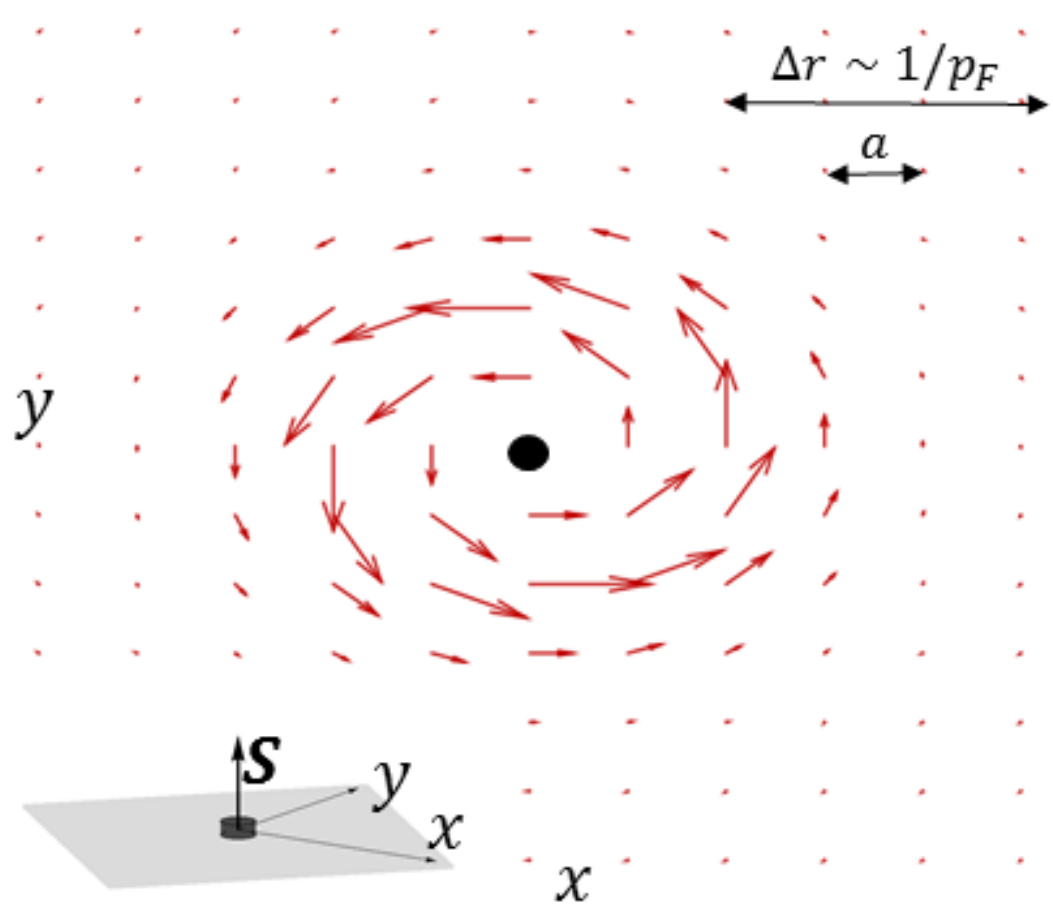}  \\
\vspace{0.4in}
(b) \includegraphics[width=0.7\linewidth]{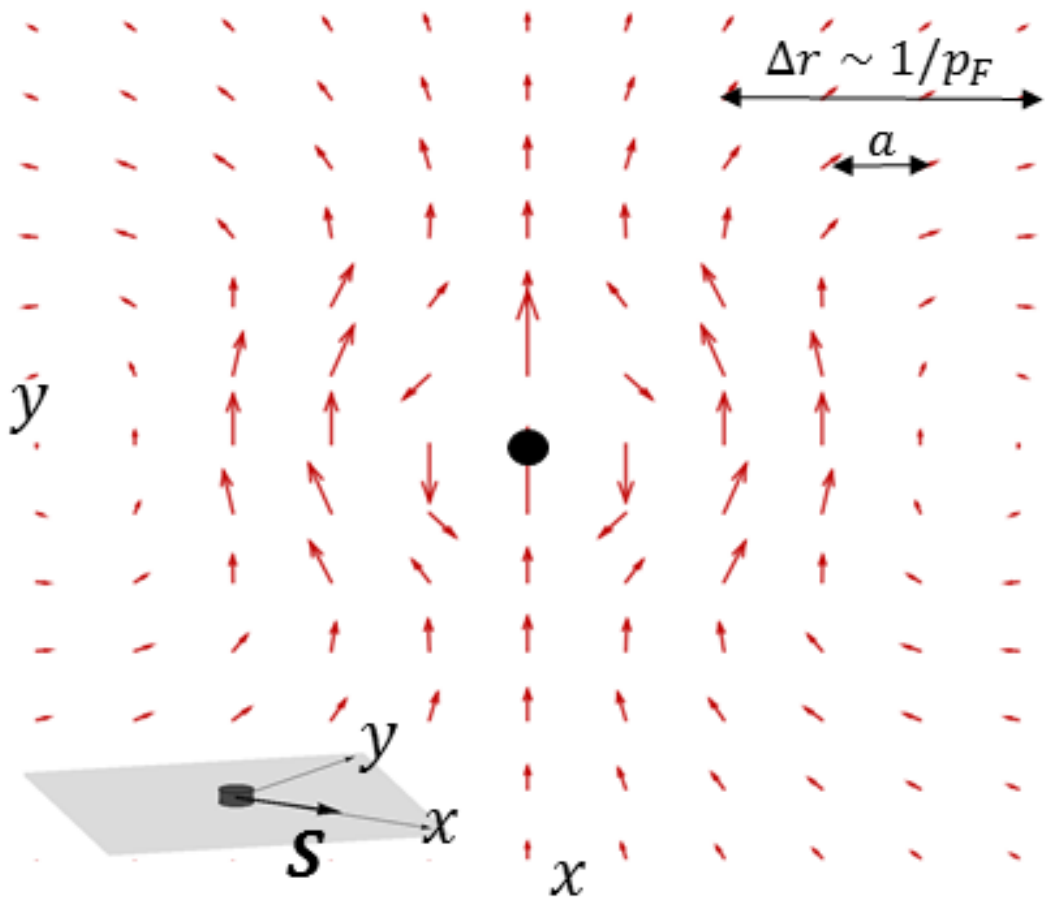}
\caption{(Color online.) Current around a point magnetic impurity in a square tight-binding model obtained using self-consistent calculations. The lattice constant is $a$ and the current vector is indicated on each tight-binding site with the impurity marked with a black dot. The direction of the impurity magnetization is shown in the insets.} \label{fig4}
\end{figure}
%%%%%%%%%%%%%%%%%%%%%%%%%%%%%%%%%%%%%%%%%%%%%%%%%%%%%%%%%%%%%%%%%%%%%%%%%
\subsection{Green's function in the real space}
%%%%%%%%%%%%%%%%%%%%%%%%%%%%%%%%%%%%%%%%%%%%%%%%%%%%%%%%%%%%%%%%%%%%%%%%%
We calculate the Green's function in the real space as
\begin{align}
	 g_{\bm r 0}(\omega) = \int\limits_{\bm p}\,\frac{e^{i\,\bm p\cdot\bm r}}{\omega-[\xi(\bm p)+\lambda(\bm\sigma\times\bm p)_z]\tau_z-\Delta\tau_x} 
	\label{agr}
\end{align}
We substitute the expansion of the Green's function (\ref{gExp}) in Eq.~(\ref{agr}) 
\begin{align*}
         g_{\bm r 0}(\omega) =\frac{1}{2}\int\limits_{\bm p}\,&\left[\frac{1+i(\bm \sigma\times \hat{\bm p})}{\omega-\xi_+(\bm p)\tau_z-\Delta\tau_x} \right. \\ 
	            & +\left.\frac{1-i(\bm \sigma\times \hat{\bm p})}{\omega-\xi_-(\bm p)\tau_z-\Delta\tau_x}\right], 
\end{align*}
where $\xi_\pm (\bm p) = \xi(\bm p)\pm \lambda p$. The angular integration in the momentum space transforms the Rashba term in the momentum space $(\bm \sigma\times \hat{\bm p})$ into a corresponding term in real space and produces the Bessel functions $J_0$ and $J_1$
\begin{align*}
                g_{\bm r 0}(\omega)=\frac{1}{2}\int\limits_{\bm p}\,&\left[\frac{J_0(p r)+i(\bm \sigma\times \hat{\bm r})\,J_1(pr)}{\omega-\xi_+(\bm p)\tau_z-\Delta\tau_x} \right. \\ 
	            & +\left.\frac{J_0(p r)-i(\bm \sigma\times \hat{\bm r})\,J_1(pr)}{\omega-\xi_-(\bm p)\tau_z-\Delta\tau_x}\right], 
\end{align*}
For small distance $r\ll\xi$, we can substitute the momenta in the Bessel function with their average values $p =p^\pm_F$, which satisfy equations $\xi_\pm \left( p_F^{\pm} \right) = 0$. Then numerators do not depend on the integration variable $\bm p$, and, thus, integration gives
\begin{align}
	 g_{\bm r 0}(\omega)  &=-\pi\frac{\omega+\Delta\tau_x}{\sqrt{\Delta^2-\omega^2}}  \left[f_0(r)+i(\bm \sigma\times \bm r)\,f_1(r)\right],\label{grf} \\ 
	& {\rm where}\,\,\, f_0(r) = \frac{1}{2}\left[\rho_+J_0(p_F^+ r)+\rho_-J_0(p_F^- r)\right],\nonumber \\
	& \qquad\,\,\,\,\, f_1(r) = \frac{1}{2r}\left[\rho_+J_1(p_F^+r) -\rho_-J_1(p_F^-r)\right]. \nonumber
\end{align}
In Eq.~(\ref{grf}), the density of states for each branch of the Rashba spectrum has been defined as
\begin{equation}
	\rho_\pm = \frac{\rho_0}{2}\, \frac{1}{1\pm m\lambda/p_F^\pm},\quad \rho_0 = \frac{m}{\pi}.
\end{equation}
For example at $r=0$, the Green's function has a conventional form
\begin{equation}
             g_{0 0}(\omega) = -\frac{\pi\rho}{2}\,\frac{\omega+\Delta\tau_x}{\sqrt{\Delta^2-\omega^2}}, 
	\label{g0}
\end{equation}
where $\rho = \rho_++\rho_-$. 
%%%%%%%%%%%%%%%%%%%%%%%%%%%%%%%%%%%%%%%%%%%%%%%%%%%%%%%%%%%%%%%%%%%%%%%%%
\subsection{T-matrix and spin polarization}
%%%%%%%%%%%%%%%%%%%%%%%%%%%%%%%%%%%%%%%%%%%%%%%%%%%%%%%%%%%%%%%%%%%%%%%%%

Using Eq.~(\ref{g0}), we can expand the T-matrix in Eq.~(\ref{GF})
\begin{equation*}
	T(\omega) = \frac{-\bm S \cdot \bm \sigma}{1 -\frac{\pi\rho \,\bm S \cdot \bm \sigma}{2}\,\frac{\omega+\Delta\tau_x}{\sqrt{\Delta^2-\omega^2}}},
\end{equation*}
which after insertion of projectors $(1\pm \tau_x)/2$ becomes
\begin{equation*}
	\sum\limits_{s = \pm } \frac{1+s\tau_x}{2}\, \frac{-\bm S \cdot \bm \sigma}{1 -\frac{\pi\rho \,\bm S \cdot \bm \sigma}{2}\,\frac{\omega+s\Delta}{\sqrt{\Delta^2-\omega^2}}}.
\end{equation*}
We multiply the numerator and denominator of the fraction by $1 +\frac{\pi\rho \,\bm S \cdot \bm \sigma}{2}\,\frac{\omega+s\Delta}{\sqrt{\Delta^2-\omega^2}}$ to eliminate the $\sigma$ matrices in the denominator
\begin{equation*}
	\sum\limits_{s = \pm } \frac{1+s\tau_x}{2}\, \frac{-\bm S \cdot \bm \sigma}{1 -\frac{(\pi\rho S)^2}{4}\,\frac{(\omega+s\Delta)^2}{\Delta^2-\omega^2}}\left( 1+\frac{\pi\rho \,\bm S \cdot \bm \sigma}{2}\,\frac{\omega+s\Delta}{\sqrt{\Delta^2-\omega^2}} \right)
\end{equation*}
and after a few transformations rewrite the expression in the final form
\begin{align}
	T(\omega) =  \sum \limits_{s = \pm} &\frac{1+s \tau_x}{2}\,  \frac{s \Delta-\omega}{\left[ 1 + \frac{(\pi\rho S)^2}{4} \right](\omega-E^s_{\rm YSR})} \nonumber \\
	& \times \left( \bm S \cdot \bm \sigma+\frac{\pi\rho S^2}{2}\,\frac{\omega+s\Delta}{\sqrt{\Delta^2-\omega^2}} \right), \label{tm}
\end{align}
where the YSR energies $E^{\pm}_{\rm YSR}$ are given in Eq.~(\ref{Shiba}). We substitute Eq.~(\ref{tm}) in the expression for the current~(\ref{curGF}) and find that the last term in the parenthesis on the second line vanishes after taking the trace. The remaining term proportional to $\bm S \cdot \bm \sigma$ has a pole at the YSR energy and gives a non-zero contribution to the current. Note that, in general, we expect the contribution to the current both from the localized subgap states and the delocalized supragap states. 

Using the expansion of the T-matrix in Eq.~(\ref{tm}), we calculate the spin-polarized LDOS~(\ref{spldos}) in the vicinity of the positive YSR state
\begin{align}
	\rho^{\pm x}_{r\hat{\bm x}}(\omega) =\left[ f_0(r) \pm r f_1(r)\right]^2 \,\frac{S\Delta\,\delta(\omega-E^+_{\rm YSR})}{2\left[ 1+\left( \frac{\pi\rho S}{2} \right)^2 \right]^2}\, \label{sppol}
\end{align}
for, e.g., perpendicular local moment $\bm S = S \hat{\bm z}$. The first term in square brackets is responsible for distinct SP-LDOS in the opposite directions $\pm \hat{\bm x}$, whereas the second fraction in Eq.~(\ref{sppol}) determines the overall strength of the YSR state. The terms $f_0(r)$ and $rf_1(r)$, defined in Eq.~(\ref{grf}), are of the same order sufficiently far from the impurity, and the YSR state acquires strong in-plane spin polarization $\sigma^{x}_{r\hat{\bm x}} = \rho^{+x}_{r\hat{\bm x}}-\rho^{-x}_{r\hat{\bm x}} \approx  \rho^{+x}_{r\hat{\bm x}}$ according to Eq.~(\ref{sppol}). Such a large spin polarization should be possible to detect experimentally using SP-STM. 
%%%%%%%%%%%%%%%%%%%%%%%%%%%%%%%%%%%%%%%%%%%%%%%%%%%%%%%%%%%%%%%%%%%%%%%%%%%%
\section{Numerical simulation} \label{sec:numerical}
%%%%%%%%%%%%%%%%%%%%%%%%%%%%%%%%%%%%%%%%%%%%%%%%%%%%%%%%%%%%%%%%%%%%%%%%%%%%i
We consider the following tight-binding model on a square lattice (see e.g. Ref.~\cite{Kristofer2013,Kristofer2015}):
\begin{align}
	H &= H_{\rm kin} + {H}_{\rm so} + {H}_{\rm sc}+ {H}_{\rm imp},  \label{fullEq} \\ 
	&H_{\rm kin} = -t\sum_{\langle\bm{i},\bm{j}\rangle,\sigma}c_{\bm{i}\sigma}^{\dagger}c_{\bm{j}\sigma} - \mu\sum_{\bm{i},\sigma}c_{\bm{i}\sigma}^{\dagger}c_{\bm{i}\sigma}, \nonumber \\ 
	&H_{\rm so} = \frac{i\lambda}{2}\sum_{\bm{i}\bm{b}}\,c_{\bm{i}+\bm{b}\sigma}^{\dagger}\hat{\bm z}\cdot(\bm\sigma\times\bm b)_{\sigma\sigma'}\,c_{\bm i\sigma'} + {\rm H.c.}, \nonumber \\ 
	&H_{\rm sc} = \sum_{\bm{i}}\Delta_{\bm{i}}c_{\bm{i}\uparrow}^{\dagger}c_{\bm{i}\downarrow}^{\dagger} + {\rm H.c.} \nonumber. \\
	&H_{\rm imp} = -c_{0\sigma}^{\dagger}\left( \bm S\cdot \bm\sigma\right)_{\sigma\sigma'}c_{0\sigma'}, \nonumber  
\end{align}
where $\bm i$ and $\bm j$ are site indices and $\bm b$ is a unit vector pointing along one of the four types of bonds on the square lattice. The parameters of the model are the strength of the nearest neighbor hopping $t$, the chemical potential $\mu $, and the ferromagnetic vector $\bm S$. Also, a superconducting pair potential $v_{\rm sc}$ is used to self-consistently determine the superconducting order parameter through
\begin{align}
	\Delta_{\bm{i}} =& -v_{\rm sc}\langle c_{\bm{i}\downarrow}c_{\bm{i}\uparrow}\rangle\nonumber\\
	=& -v_{sc}\sum_{E_{\nu}<0}v_{\nu\bm{i}\downarrow}^{*}u_{\bm{i}\uparrow}. \label{selfCons}
\end{align}
We solve Eqs.~(\ref{fullEq}) and (\ref{selfCons}) self-consistently and calculate currents around  point magnetic impurities using the expressions derived below in Appendix \ref{sec:numerical_currents}. Figure~\ref{fig3} corresponds to the following parameters: $t = 1$, $\mu = -4$, $\lambda = 0.56$, and $v_{\rm sc} = 5.36$. These values give a generic band structure of a lightly hole-doped Rashba SOC semiconductor in proximity to a conventional $s$-wave superconductor.
Panels (b), (c) and (d) correspond to a magnetic impurity with $S=1.6, 2.72$, and $7.36$ respectively. Figure 4 is plotted for the parameters $t = 1$, $\mu = -4$, $\lambda = 0.56$, $v_{\rm sc} = 5.36$, and $S=2.56$, which is just below $S_c$.
%%%%%%%%%%%%%%%%%%%%%%%%%%%%%%%%%%%%%%%%%%%%%%%%%%%%%%%%%%%%%%%%%%%%%%%%%%%%
\subsection{Calculating current} \label{sec:numerical_currents}
%%%%%%%%%%%%%%%%%%%%%%%%%%%%%%%%%%%%%%%%%%%%%%%%%%%%%%%%%%%%%%%%%%%%%%%%%%%%
Numerical expressions for the current are derived by considering the time rate of change of the density operator $\hat{\rho}_{\mathbf{i}} = \sum_{\mathbf{i}\sigma}c_{\mathbf{i}\sigma}^{\dagger}c_{\mathbf{i}\sigma}$ (see e.g.~Ref.~\cite{Black-Schaffer2008}):
\begin{align}
	\label{Equation:Time_rate_of_change_density}
	\frac{d\hat{\rho}_{\mathbf{i}}}{dt} =& \frac{i}{\hbar}\left[\mathcal{H}, \hat{\rho}_{\mathbf{i}}\right].
\end{align}
Let $\sigma$ and $\bar{\sigma}$ be opposite spins and define
\begin{align}
	\label{Equation:Current_operators}
	\hat{S}_{\sigma}^{\mathbf{i}} =& a_{\bar{\sigma}\sigma}^{\mathbf{i}}c_{\mathbf{i}\sigma}^{\dagger}c_{\mathbf{i}\bar{\sigma}} + a_{\bar{\sigma}\sigma}^{\mathbf{i}*}c_{\mathbf{i}\bar{\sigma}}^{\dagger}c_{\mathbf{i}\sigma},\nonumber\\
	\hat{J}_{\sigma\sigma}^{\mathbf{i}\mathbf{b}} =& - \left(a_{\sigma\sigma}^{\mathbf{i}\mathbf{b}}c_{\mathbf{i}+\mathbf{b}\sigma}^{\dagger}c_{\mathbf{i}\sigma} + a_{\sigma\sigma}^{\mathbf{i}\mathbf{b}*}c_{\mathbf{i}\sigma}^{\dagger}c_{\mathbf{i}+\mathbf{b}\sigma}\right),\nonumber\\
	\hat{J}_{\sigma\bar{\sigma}}^{\mathbf{i}\mathbf{b}} =& -\left(a_{\bar{\sigma}\sigma}^{\mathbf{i}\mathbf{b}}c_{\mathbf{i}+\mathbf{b}\bar{\sigma}}^{\dagger}c_{\mathbf{i}\sigma} + a_{\bar{\sigma}\sigma}^{\mathbf{i}\mathbf{b}*}c_{\mathbf{i}\sigma}^{\dagger}c_{\mathbf{i}+\mathbf{b}\bar{\sigma}}\right),
\end{align}
where $\mathbf{b}$ runs over the vectors from site $\mathbf{i}$ to its nearest neighbors and $a_{\bar{\sigma}\sigma}^{\mathbf{i}}, a_{\sigma\sigma}^{\mathbf{i}\mathbf{b}}$, and $a_{\bar{\sigma}\sigma}^{\mathbf{i}\mathbf{b}}$ are coefficients to be determined.
It can be shown that Eq.~\eqref{Equation:Time_rate_of_change_density} can be written as
\begin{align}
	\frac{d\hat{\rho}_{\mathbf{i}}}{dt} =& \sum_{\sigma}\hat{S}_{\sigma}^{\mathbf{i}} - \sum_{\mathbf{b}\sigma}\left(\hat{J}_{\sigma\sigma}^{\mathbf{i}\mathbf{b}} + J_{\sigma\bar{\sigma}}^{\mathbf{i}\mathbf{b}}\right).
\end{align}
It is clear from the definition of Eq.~\eqref{Equation:Current_operators} that $S_{\sigma}^{\mathbf{i}}$ is an on-site source operator which converts $\bar{\sigma}$ spins into $\sigma$, while $J_{\sigma\sigma}^{\mathbf{i}\mathbf{b}}$ is a current operator for $\sigma$-spins away from site $\mathbf{i}$ along bond $\mathbf{b}$.
Similarly $J_{\sigma\bar{\sigma}}^{\mathbf{i}\mathbf{b}}$ is a current operator for $\sigma$-spins away from site $\mathbf{i}$ along bond $\mathbf{b}$, but for spins which are flipped in the transfer process. 
Summing the two types of currents for both spin species, we arrive at the expression
\begin{align}
	\hat{J}^{\mathbf{i}\mathbf{b}} =& \sum_{\sigma\sigma'}\hat{J}_{\sigma\sigma'}^{\mathbf{i}\mathbf{b}},
\end{align}
for the total operator for currents away from site $\mathbf{i}$ along bond $\mathbf{b}$. The current operator on site $\mathbf{i}$ can now be defined as
\begin{align}
	\label{Equation:Current_operator_on_site}
	\hat{\mathbf{J}}^{\mathbf{i}} =& \sum_{\mathbf{b}}\mathbf{b}\hat{J}^{\mathbf{i}\mathbf{b}}.
\end{align}
Evaluation of the commutator in Eq.~\eqref{Equation:Time_rate_of_change_density} reveals that the relevant coefficients in Eq.~\eqref{Equation:Current_operators} are
\begin{align}
	\label{Equation:Current_coefficients}
%	a_{\uparrow\uparrow}^{\mathbf{i}\mathbf{b}} = %a_{\downarrow\downarrow}^{\mathbf{i}\mathbf{b}} =& -it,\nonumber\\
%	a_{\uparrow\downarrow}^{\mathbf{i}\mathbf{b}} =& -ie^{-%i\theta_{\mathbf{b}}}\frac{\lambda}{2},\nonumber\\
%	a_{\downarrow\uparrow}^{\mathbf{i}\mathbf{b}} =& %ie^{i\theta_{\mathbf{b}}}\frac{\lambda}{2}.
		a_{\sigma\sigma}^{\mathbf{i}\mathbf{b}} =& -it,\nonumber\\
	a_{\bar{\sigma}\sigma}^{\mathbf{i}\mathbf{b}} =& -\hat{\bm z}\cdot\left(\bm\sigma\times\bm b\right)_{\bar{\sigma}\sigma}\frac{\lambda}{2}.
\end{align}

Using Eq.~\eqref{Equation:Current_operators}, \eqref{Equation:Current_operator_on_site}, and \eqref{Equation:Current_coefficients} the current can finally be calculated as
\begin{align}
	\left\langle \hat{\mathbf{J}}^{\mathbf{i}}\right\rangle = \sum_{\mathbf{b}\sigma,E_{\nu}<0}&\mathbf{b}\left\{it
	\left(v_{\mathbf{i}+\mathbf{b}\sigma}^{*}u_{\mathbf{i}\sigma} - v_{\mathbf{i}\sigma}^{*}u_{\mathbf{i}+\mathbf{b}\sigma}\right)\right.\nonumber\\
	&+ \left.\frac{\lambda}{2}\left[\hat{\bm z}\cdot\left(\bm\sigma\times\bm b\right)_{\bar{\sigma}\sigma}v_{\mathbf{i}+\mathbf{b}\bar{\sigma}}^{*}
	u_{\mathbf{i}\sigma}\right.\right.\nonumber\\
	&\;\;\;\;\;+ \left.\left.\hat{\bm z}\cdot\left(\bm\sigma\times\bm b\right)_{\bar{\sigma}\sigma}^{*}v_{\mathbf{i}\sigma}^{*}
u_{\mathbf{i}+\mathbf{b}\bar{\sigma}}\right]\right\}.
\end{align}

\end{document}